\documentclass[aps, pra, a4paper, showpacs, twocolumn, english, 10pt, nofootinbib]{revtex4-2}
\usepackage[T1]{fontenc}
\usepackage{babel}
\usepackage{bbm, amsthm, bm, textcomp, nicefrac, geometry, ragged2e}
\usepackage[dvipsnames]{xcolor}
\usepackage{float}
\usepackage[bbgreekl]{mathbbol}
\usepackage{color, verbatim, enumitem, ulem}
\usepackage{pbox}
\usepackage{mathrsfs}
\usepackage{physics}
\usepackage{amssymb}
\usepackage{mathtools}
\usepackage{stackrel}
\usepackage[thinlines]{easytable}
\usepackage{amsmath}
\usepackage{makecell}
\usepackage{aligned-overset}

\mathtoolsset{showonlyrefs}

\usepackage{array}
\usepackage{multirow}

\usepackage{graphicx}
\usepackage[caption=false]{subfig}
\addto\captionsenglish{}

\usepackage{hyperref}
\hypersetup{
    colorlinks = true,
    linkcolor = BrickRed,
    citecolor = BrickRed,
    filecolor = Black,     
    urlcolor = Black,
}

\newcommand{\Fd}{\mathbb{F}_d}
\newcommand{\Fp}{\mathbb{F}_p}
\newcommand{\comrel}{\mathrel{\overset{\text{c.r.}}{=}}}
\newcommand{\uprule}{\mathrel{\overset{\text{u.r.}}{=}}}
\newcommand{\Fdn}{\mathbb{F}^n_d}

\DeclareMathSymbol{\shortminus}{\mathbin}{AMSa}{"39}
\newcommand{\sm}{\shortminus}

\renewcommand\bra[1]{{\langle{#1}|}}
\renewcommand\ket[1]{{|{#1}\rangle}}

\begin{document}

\title{Qudit Noisy Stabilizer Formalism}
\author{Paul Aigner, Maria Flors Mor-Ruiz, and Wolfgang D\"ur}
\affiliation{Institut f\"ur Theoretische Physik, Universit\"at Innsbruck, Technikerstra{\ss}e 21a, 6020 Innsbruck, Austria}

\begin{abstract}
We introduce the qudit noisy stabilizer formalism, a framework for efficiently describing the evolution of stabilizer states in prime-power dimensions subject to generalized Pauli-diagonal noise under Clifford operations and generalized Pauli measurements. For arbitrary dimensions, the formalism remains applicable, though restricted to a subset of stabilizer states and operations. The computational complexity scales linearly with the number of qudits in the initial state and exponentially with the number of qudits in the final state. This ensures that when noisy qudit stabilizer states evolve via generalized Pauli measurements and Clifford operations to generate multipartite entangled states of a few qudits, their description remains efficient. We demonstrate this by analyzing the generation of a generalized Bell pair from a noisy linear cluster state subject to two distinct noise sources acting on each of the qudits.
\end{abstract}

\maketitle

\section{Introduction} \label{Sec.Introduction}
High-dimensional quantum systems, known as qudits, have been identified as promising candidates for quantum information processing, offering distinct advantages over the conventional qubit. These advantages include enhanced performance in fault-tolerant quantum computation~\cite{PhysRevA.92.032309,PhysRevA.91.042331,PhysRevA.71.022316}, quantum communication tasks such as quantum key distribution~\cite{PhysRevA.64.012306, Miller_2018, PhysRevA.73.032325, Doda_QKD}, quantum steering~\cite{Srivastav2022}, and entanglement purification~\cite{Miguel_Purification, PhysRevA.72.032313}, as well as improvements in quantum metrology~\cite{PhysRevA.97.022115} and quantum simulation~\cite{PhysRevA.110.062602}. Besides the potential these systems pose, their study is further motivated due to the fact that they naturally arise as multilevel energy systems in several platforms, e.g., trapped ions \cite{ringbauer2022universal, hrmo2023native, Low_2020}, superconducting qudits \cite{PhysRevX.13.021028,doi:10.1126/science.1173440}, neutral atoms \cite{PRXQuantum.4.040333,PhysRevApplied.19.034089}, and they can be even realized in information carriers such as photons \cite{PhysRevX.5.041017, Zhong_2015, doi:10.1126/sciadv.1701491}.

Given the relevance of these high-dimensional systems, their classical simulation is key for exploring the potential they hold. In general, this simulation is considered to be hard due to the exponential growth of the Hilbert space with respect to the system size. Nevertheless, for some restricted subclasses of states, efficient classical simulation is possible. Of particular note are stabilizer states, uniquely defined by linearly many stabilizer operators, whose evolution under Clifford operations and Pauli measurements can be efficiently described by updating these operators, as guaranteed by the (generalized) Gottesman-Knill theorem~\cite{2-2013, osti_319738, PhysRevA.70.052328, van2010classical}. Besides their computational efficiency, stabilizer states and graph states, which are local Clifford equivalent to stabilizer states \cite{grassl2002graphs, schlingemann2001stabilizercodesrealizedgraph}, are widely used as the foundation for various quantum protocols, e.g., in quantum networks \cite{Englbrecht2022transformationsof, Keet2010,mansour2020quantum,colrecurrence}, in quantum error correction \cite{9321171,PhysRevA.103.042420,PhysRevA.84.052306,PhysRevA.105.032450}, and for measurement-based quantum computation of Clifford circuits \cite{PhysRevA.68.062303,PhysRevA.95.032312}.

Physical systems are inherently affected by noise, imperfections, and decoherence, thus, the analysis of protocols should incorporate these conditions. Here, a mixed-state description is required, leading to a more complicated and less efficient description of the states and processes acting on them. For stabilizer states, most of the existing approaches for handling noise rely on the efficient classical simulation of Clifford circuits \cite{PhysRevA.99.052307, brandl2024efficient}, where noise processes are addressed through Monte Carlo sampling via the random insertion of additional gates. Nevertheless, this approach does not yield the full noisy state but merely enables sampling from the output distribution. Another prominent strategy involves encapsulating all individual noise contributions within an error probability tensor and propagating the noisy circuit by updating this tensor \cite{Miller_2018}. While this method provides a systematic framework for analyzing noise, the size of the error probability tensor grows exponentially with the number of noise channels it incorporates, rendering the approach computationally infeasible for systems with a large number of noise contributions.

Here we present the qudit noisy stabilizer formalism, following the two-dimensional counterpart \cite{1-2023}, providing a framework for describing prime-power dimensional stabilizer states affected by diagonal noise in the generalized Pauli basis. The key idea of this formalism is to treat noise channels and quantum states independently, eliminating the need to explicitly handle density matrices. Using this separation, we establish rules to update error channels, along with efficiently updating stabilizer states under Clifford operations and generalized Pauli measurements. The update rules for noise channels are derived specifically for graph states but can be generalized to stabilizer states due to the local Clifford equivalence between these two classes of states. As a result, this formalism provides a computationally efficient and analytical approach, scaling linearly with the number of noise channels and initial subsystems. Nevertheless, it scales exponentially with respect to the number of subsystems present at the end of the protocol.

Consequently, the method is particularly well suited for algorithms involving small target states, such as the generation of generalized Bell states in quantum communication scenarios. To illustrate its practical utility, we investigate the generation of generalized Bell pairs from a linear cluster state under two noise sources, highlighting the analytical capabilities offered by the qudit noisy stabilizer formalism. Similar scenarios in the framework of quantum communication and quantum repeaters have also been tackled in the qubit case \cite{Mor_Influence, MorRuiz2025imperfectquantum, morruiz2025mergingbasedquantumrepeater}, successfully employing the noisy stabilizer formalism for qubits \cite{1-2023} and demonstrating its practical utility in describing and analyzing noisy quantum systems.

The structure of the paper is as follows. In Sec.~\ref{Sec.Background}, the necessary background for qudits together with the qubit noisy stabilizer formalism are reviewed. Next, in Sec.~\ref{Sec.Weyl measurements of graph states}, the effect of Weyl measurements on graph states in prime-power dimension is demonstrated. This is followed, in Sec.~\ref{Sec.NSF}, by the description of the method to efficiently describe the manipulation of noisy prime-power dimensional graph states, whose efficiency is examined in Sec.~\ref{Sec.Efficiency}. In Sec.~\ref{Sec.Generalizations}, the discussion on the extension of the presented method to stabilizer states and to arbitrary dimension is presented, as well as the analysis of the limitation to Pauli-diagonal noise models. Then, in Sec.~\ref{Sec.Application of method}, we focus on the application of the formalism to the analysis of the generation of a generalized Bell-pair from a noisy linear cluster state subject to two distinct noise sources acting on each of the qudits, thereby showcasing its utility. Finally, in Sec.~\ref{Sec.Conclusions and outlook}, we summarize the results and offer a concise overview of the potential applications of the formalism.

\section{Background}
\label{Sec.Background}
Here we present the mathematical background for odd prime-power dimensions \cite{1-2021,gross2006hudson}. The construction for the even case is given in Appendix~\ref{Sec. Background for even prime power dimension}.

\subsection{Generalized Pauli group} \label{Subsec.The generalized Pauli group}
In prime-power dimension $d=p^m$, where $p$ is a prime and $m$ is a positive integer, one can represent linear operators with generalized Pauli operators. The generalized Pauli $X$ operator is defined as 
\begin{equation}\label{eq:pauli:x}
    X(x):=\sum_{y \in \Fd} \ket{x+y}\bra{y},
\end{equation}
where $x$ is an element of $\Fd$, the unique finite field with $d=p^m$ elements. The elements $x\in \Fd$ are given as polynomials of the form
\begin{equation}
x=x_0+x_1\theta+x_2\theta^2+\cdots+ x_{m \sm 1}\theta^{m \sm 1},
\end{equation}
where the coefficient $x_i \in \Fp$ and $\theta$ is the root of an irreducible polynomial with degree $m$ over $\Fp$. An irreducible polynomial over $\Fp$ is a polynomial with coefficients in $\Fp$ that cannot be written as a product of two nonconstant polynomials with coefficients in $\Fp$. Elements in $\Fd$ are added via the addition of their coefficients, while the multiplication is defined as the multiplication of polynomials modulo the irreducible polynomial. The generalized Pauli $Z$ operator is defined as 
\begin{equation}\label{eq:pauli:z}
    Z(z):=\sum_{y \in \Fd} \chi(zy) \ketbra{y}, 
\end{equation}
where $z \in \Fd$ and $\chi(zy):=\omega^{\text{tr}(zy)}$, with $\omega=e^{\frac{2 \pi i}{p}}$ and
\begin{equation}
    \text{tr}(x):= \sum_{n=0}^{m \sm 1} x^{p^n}
\end{equation}
being the finite field trace mapping polynomial elements of $\Fd$ to integer elements of $\Fp$. 

The generalized Pauli group is defined as the Heisenberg-Weyl group
\begin{equation}
    \text{HW}:=\{W(z,x,t), \ z,x,t \in \Fd \},
\end{equation}
where 
\begin{equation}
W(z,x,t):= \chi(t)\chi(\sm 2^{\sm 1}z x)Z(z)X(x)
\end{equation}
is the Weyl operator. The generalized phase is represented by $\chi(t)$, and for $t=0$ we set $W(z,x):=W(z,x,0)$.

The $n$-qudit Weyl operators are defined as tensor products of the single-qudit Weyl operators, and are written as
\begin{equation}
    W(\bm z,\bm x,t):= \chi(t)\chi(\sm 2^{\sm 1}\bm z \bm x)Z(\bm z)X(\bm x),
\end{equation}
where $\bm x \in \Fdn$ and $O(\bm x)=\bigotimes_{i}O_i(\bm x_i)$. Therefore, the $n$-qudit Heisenberg-Weyl group is defined as
\begin{equation}
    \text{HW}_n:=\{W(\bm z,\bm x,t), \ \bm z,\bm x\in \Fdn, \ t \in \Fd\}.
\end{equation}
\subsection{Stabilizer formalism} \label{Subsec.The Stabilizer formalism}
Stabilizer groups are Abelian (commutative) subgroups of the Heisenberg-Weyl group. They can be described via isotropic subspaces $K \subset \Fd^{2n}$. These are subsets of their symplectic complement
\begin{equation}\label{eq:isotropic}
K^\perp:=\{\bm v=(\bm z, \bm x) \in \Fd^{2n}, | \, [\bm v,\bm w]=0 \ \forall \bm w \in K \},
\end{equation}
where $[\bm v,\bm w]$ is the symplectic product defined by
\begin{equation}
    [\bm v,\bm w]:= \bm v^TJ \bm w, \quad J=\begin{pmatrix}
0 & \mathbb{1} \\
\sm \mathbb{1} & 0 
\end{pmatrix}.
\end{equation}

If the isotropic subspace is maximal, then it has dimension $\text{dim}(\Fd^{2n})/2$ and is called Lagrangian and denoted by $\mathbbm{L}$. A Lagrangian $\mathbbm{L}$ uniquely defines a so-called stabilizer state by the following eigenvalue equations
\begin{equation}  \label{Eq.Stabilizer_eigen}
    \chi([\bm v,\bm w])W(\bm w)\ket{\mathbbm{L},\bm v}= \ket{\mathbbm{L},\bm v} \ \forall \bm w \in \mathbbm{L}.
\end{equation}
We call the operators fulfilling Eq.~\eqref{Eq.Stabilizer_eigen}, stabilizer operators, and they constitute a group 
\begin{equation}
    \mathcal{S}=\{\chi([\bm v,\bm w])W(\bm w), \ \bm w \in \mathbbm{L} \},
\end{equation}
which we call the stabilizer (group) of $\ket{\mathbbm{L},\bm v}$. We can specify a stabilizer state by specifying the Lagrangian subspace $\mathbbm{L}$ associated to it. This is done by fixing a basis $\{\bm v_1,\bm v_2, \dots, \bm v_n \}$ of $\mathbbm{L}$. Thus, instead of representing a stabilizer state by an exponentially large state vector of size $\mathcal{O}(d^n)$, we can uniquely specify it by $n$ stabilizers and a phase vector.

\subsection{Clifford group} \label{Subsec.The Clifford group}
The unitary operators, which map stabilizer states to stabilizer states, are called Clifford operators. They form a group $\text{Cl}_n$, which is generated by the set of operators given in Table \ref{Table:1}. Moreover, the local Clifford group, $\text{Cl}_1$, is generated by the Clifford generators of Table \ref{Table:1} without the $CZ$ operator.
The Clifford group $\text{Cl}_n$ normalizes the Heisenberg-Weyl group
\begin{equation}
    U \text{HW}_n U^\dag = \text{HW}_n, \ \forall U \in \text{Cl}_n,
\end{equation}
such that Clifford and Weyl operators commute up to a change of the Weyl operator.
\begin{table}[h!]
\begin{equation*}
\begin{aligned}[c]
Z\ket{x} &=  \chi(x) \ket{x}\\
X \ket{x}&= \ket{x+1} \\
H\ket{x}&= \ket{\hat x}
\end{aligned}
\quad \quad
\begin{aligned}[c]
CZ \ket{x,y}&= \chi(xy) \ket{x,y} \\
 S(\lambda) \ket{x}&=  \chi\left(\lambda x^2/2\right) \ket{x}\\ M(\lambda)\ket{x}
&= \ket{\lambda x}
\end{aligned}
\end{equation*}
\caption{List of generators of the Clifford group, with $\lambda \in \Fd$ and $\ket{\hat x}:=\frac{1}{\sqrt{d}} \sum_{y \in \Fd } \chi(xy)\ket{y}$.} \label{Table:1}
\end{table}

The Heisenberg-Weyl group $\text{HW}_n$ and the Weil representation of the symplectic group $\text{Sp}_n$, generated by the Clifford generators without the $X$ and $Z$ operators, form the Clifford group $\text{Cl}_n$.
Elements of $U_g \in \text{Sp}_n$ map Weyl operators $W(\bm v),W(\bm  w)$ with $[\bm v, \bm w]=0$ under conjugation to Weyl operators $W( g(\bm v)),W(  g(\bm w))$ with $[ g(\bm v),  g(\bm w)]=0$ and $g$ being an automorphism on $\Fd^{2n}$. Therefore, the conjugation leaves the symplectic product invariant. Conjugation of a Weyl operator $W(\bm v)$ by a Weyl operator $W(\bm w)$ leads to $\chi([\bm v,\bm w]) W(\bm v)$. Consider the action of a Clifford unitary $U=S_g W(\bm q)$ on a stabilizer state $\ket{\mathbbm{L},\bm v}$, 
\begin{equation}
    \begin{aligned}
    U \ket{\mathbbm{L},\bm v}&= U \chi([\bm v,\bm w])W(\bm w)U^\dag U \ket{\mathbbm{L},\bm v} \\
    &=\chi([\bm v+ \bm q,\bm w]) W(\bm g(w)) U\ket{\mathbbm{L},\bm v}.
\end{aligned}
\end{equation}

This implies that $U\ket{\mathbbm{L},\bm v}$ is a new stabilizer state stabilized by the stabilizer group $\{\chi([\bm v + \bm q,\bm w]) W( g(\bm w)), \bm w \in \mathbbm{L} \}=\{\chi([\bm v + \bm q, g^{\sm 1}(\bm w)]) W(\bm w), \bm w \in g(\mathbbm{L}) \}$. Therefore, we can efficiently describe the update of the stabilizer state under Clifford operations only by the action on the stabilizers, as shown by the Gottesman-Knill theorem \cite{osti_319738, PhysRevA.70.052328, van2010classical}.

\subsection{Graph states} \label{Subsec.Graph states}
\begin{figure}[h]
    \centering
    \includegraphics[width=1\linewidth]{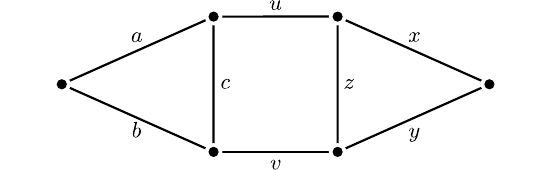}
    \caption{Weighted graph, where dots represent vertices, lines between them represent edges, and variables on the edges represent the edge weights.}
    \label{fig:example_graph}
\end{figure}

Graph states \cite{grassl2002graphs, schlingemann2001stabilizercodesrealizedgraph} are a subclass of stabilizer states. They can be described by a simple weighted graph $G=(V, E, A)$, where $V$ is the set of vertices, $E$ the set of edges, and $A$ is the adjacency matrix, such that $A_{ij} \in \Fd$, and denotes the set of weights of the edges. An instance of a simple graph is presented in Fig.~\ref{fig:example_graph}.

The state associated with a certain graph $G$, is a stabilizer state of the form $\ket{G}=\ket{\mathbbm{L},\bm 0}$ that fulfills the following eigenvalue equations
\begin{equation}\label{Eq Graph}
    X_v(x) Z(x A_v) \ket{G}= \ket{G}, \ \forall v \in V, x \in \Fd,
\end{equation}
where we use the notation $O_v(x):=O(x \bm e_v)$ with $\bm e_v$ being the $v$th standard unit vector from here onward, $A_v$ is the $v$th row of the adjacency matrix $A$. Importantly, every stabilizer state can be described by a graph state up to local Clifford operations \cite{grassl2002graphs,schlingemann2001stabilizercodesrealizedgraph}. 
\subsection{Clifford manipulation of graph states} \label{Sec.Manipulation of graph states}
\begin{figure}
    \centering
    \includegraphics[width=1\linewidth]{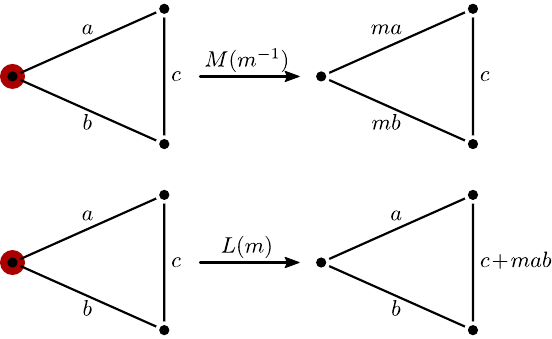}
    \caption{Illustration of local multiplication (above) and local complementation (below) on a graph state. The node where the local operations are applied is shaded.}
    \label{fig:LC_LM}
\end{figure}

The action of the local Clifford group on graph states can be described with two types of graph manipulations, the local multiplication and the local complementation, see Fig.~\ref{fig:LC_LM}. A local multiplication acting on qudit $v$ from graph $G=(V, E, A)$ with factor $m$ is denoted by $G'=G \circ_m v$ and refers to updating the adjacency matrix via $A'_{v,j}=m A_{v,j}$. The corresponding graph state is $\ket{G'}= M_v(m^{\sm 1})\ket{G}$. A local complementation is denoted by $G'=\tau_v(m)(G) $, and updates the adjacency matrix to $A'_{i,j}=A_{i,j}+mA_{v,i}A_{v,j}$. The corresponding graph state is $\ket{G'}= L_v(m)\ket{G}$, where 
\begin{equation}\label{Eq.local_compl}
L_v(m):= H_v S_v(m) H_v^\dag \bigotimes_{j } S_j(\sm m A_{v,j}^2)
\end{equation}
is a Clifford operator. Furthermore, we set $R_v(m)=H_v S_v(m) H_v^\dag$ and $S(\sm m A_{v}^2)=\bigotimes_{j } S_j(\sm m A_{v,j}^2)$. Two graph states are said to be local Clifford equivalent if there is a sequence of these two operations that maps one to the other \cite{1-2008}. 

Moreover, the action of the nonlocal Clifford generator, the $CZ$ gate (defined in Table~\ref{Table:1}), corresponds to adding $1$ to the weight of the edge connecting the two vertices that is acting on, $v$ and $w$. Such that the adjacency matrix is updated to $A'_{ij}=A_{ij}+\delta_{i,v}\delta_{j,w}+\delta_{j,v}\delta_{i,w}$. 

\subsection{Noise models} \label{Subsec.Noise model}
\subsubsection{Single-qudit Pauli noise}
The action of a single-qudit Pauli noise channel $\mathcal{E}_v$ on a qudit $v$ is described by
\begin{equation}\label{eq:single:noise}
    \mathcal{E}_v \rho=\sum_{z,x \in \Fd} \lambda_{z,x} Z_v(z)X_v(x) \rho \left(Z_v(z)X_v(x) \right)^\dag,
\end{equation}
where $\lambda_{z,x} \geq 0$, $\sum_{z,x} \lambda_{z,x}=1$, and $\rho$ is a density matrix. Note that this form can be enforced via the random application of unitary gates before and after the application of a noise channel, without changing its Pauli-diagonal elements \cite{dur2005standard}. As seen in Ref.~\cite{1-2023} for $d=2$, Pauli noise channels acting on graph states can always be written in terms of $Z$. For $d=p^m$, the action of $X$ on a graph state is as follows
\begin{equation}\label{Eq.Ztypenoise}
\begin{aligned}
    X_v(x) \ket{G}&= X_v(x) X_v( \sm x) Z(\sm xA_v) \ket{G} \\
    &= Z( \sm xA_v) \ket{G},
\end{aligned}
\end{equation}
where we use that $X_v( \sm x) Z(\sm xA_v)$ is a stabilizer operator, since it corresponds to $(X_v( x) Z(xA_v))^{p\sm 1}$. Therefore, one can replace local $X$ noise on a vertex by $Z$ noise on its neighborhood, defined by $A_v$. Therefore, Eq.~\eqref{eq:single:noise}, when applied to a graph state $\rho = \ketbra{G}$, can be rewritten as
\begin{equation}
    \mathcal{E}_v \rho = \sum_{z, x \in \Fd} \lambda_{z,x} Z_v(z)Z(\sm xA_v) \rho Z_v(\sm z)Z(xA_v).
\end{equation}
\subsubsection{Multiqudit Pauli channel}
A multiqudit Pauli channel is defined by
\begin{equation}\label{eq:general:channel}
    \mathcal{E} \rho = \sum_{\bm z ,\bm x \in \Fdn} \lambda_{\bm z, \bm x} X(\bm x) Z(\bm z) \rho \left(X(\bm x) Z(\bm z)\right)^\dag,
\end{equation}
where $\lambda_{\bm z, \bm x}\geq 0$ and $\sum_{\bm z , \bm x} \lambda_{\bm z, \bm x}=1$. If this noise channel acts on a graph state, then it can be rewritten again in terms of only $Z$-type operators, using Eq.~\eqref{Eq.Ztypenoise}. 

\subsection{Noisy stabilizer formalism for qubits} \label{Subsec.Noisy stabilizer formalism}
The noisy stabilizer formalism for qubits was introduced in Ref.~\cite{1-2023} as an efficient method to fully describe and analyze Pauli-diagonal noise acting on graph states that are manipulated by Clifford operations and Pauli measurements. This full description of the noise is possible due to the independent treatment and update of graph states and noise operators. 

The key idea of these updates resides in the commutation relations between Pauli noise operators and manipulation operators, which are referred to as \textit{update rules}. Given a manipulation operator $O$ and a Pauli noise operator $N$, which act onto the state as $ON$, they commute up to a change of the noise operator such that $ON=\tilde{N}O$. This allows us to then apply the manipulation operators directly onto the graph state, which is efficiently described thanks to the Gottesman-Knill theorem \cite{osti_319738,PhysRevA.70.052328,van2010classical}. Moreover, the noise channels are also efficiently updated independently using the update rules.

Even though graph states and noise channels are updated independently, they have to be of the same size. Such that, if the final state is small, then the application of the updated noise channels is also efficient, resulting in a method for obtaining the final mixed state without computation with large density matrices. 

\subsection{Choi-Jamio\l kowski fidelity} \label{sec:Choi-Jamiolkowski}
Channel fidelities measure how close a channel is to its ideal implementation. The Choi-Jamio\l kowski fidelity \cite{gilchrist2005distance} is a measure of channel fidelity, which can be obtained from the Choi-Jamio\l kowski isomorphism \cite{jamiolkowski1972linear}, and is such a state-independent measure, which lower bounds the fidelity of an arbitrary input state, subject to this channel \cite{dur2005standard}. 

The Choi-Jamio\l kowski isomorphism relates channels (completely positive trace-preserving maps) with states, the so-called Choi-Jamio\l kowski states. Given a channel $\mathcal{E}: A \mapsto A $ the corresponding Choi-Jamio\l kowski state $E$ is given by
\begin{equation}
    E=\mathcal{E}_A \otimes \mathrm{id}_{\bar{A}} (\ketbra{\Psi}_{A\bar{A}}),
\end{equation}
where $\bar{A}$ is a copy of the subsystem $A$ and 
\begin{equation}
\ket{\Psi}_{A\bar{A}}=\frac{1}{\sqrt{d}}\sum_{i=0}^{d \sm 1} \ket{i}_A \ket{i}_{\bar{A}}
\end{equation} 
is a maximally entangled state. The Choi-Jamio\l kowski fidelity $F_{\text{Choi}}(\mathcal{E})$ of a channel $\mathcal{E}$ is given as the fidelity, $   F(\rho,\sigma):=\text{tr}\left(\sqrt{\sqrt{\rho}\sigma\sqrt{\rho}}\right)^2$, of its corresponding Choi-Jamio\l kowski state, which is linearly related to the average channel fidelity $\bar F(\mathcal{E})$, such that 
\begin{equation}
    F_{\text{Choi}}(\mathcal{E}) = \frac{\bar F(\mathcal{E}) (d+1) \sm 1}{d},
\end{equation}
where
\begin{equation}
    \bar F(\mathcal{E}):=\int \,d\Psi  \bra{\Psi} \mathcal{E}(\ketbra{\Psi}) \ket{\Psi}.
\end{equation}

\section{Weyl measurements of graph states} \label{Sec.Weyl measurements of graph states}
We harness the graphical representation inherent to graph states by deriving the graphical rules that describe the action of the measurement of local Weyl operators on prime-power graph states. Thanks to the Gottesman-Knill theorem, measurements on graph states can be efficiently and equivalently described using their stabilizers, which may also reduce the size of the stabilizer. Here, we present said manipulations and their derivation can be found in Appendix~\ref{Sec.Proofs Measurements}. 

\begin{figure}
    \centering
    \includegraphics[width=1\linewidth]{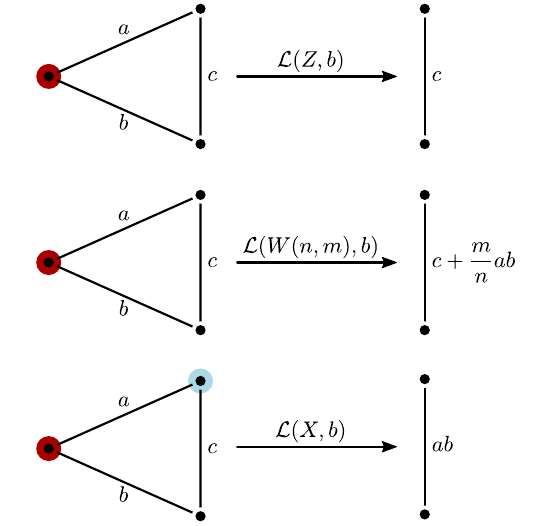}
    \caption{Illustration of Weyl measurement operations acting on a graph state. The vertex where the measurement operation is acting on is shaded dark red. The special neighbor of the $X$ measurement is shaded light blue. Note that for the $X$ measurement $a\neq 0$, since otherwise the light-blue-shaded node would not be a neighbor. }
    \label{fig:Weyl_meas}
\end{figure}

The measurement of a single-qudit Weyl operator $W_v(z,x)$ with measurement outcome $b\in \mathbbm{F}_d$ can be described by the following projector
\begin{equation}\label{Eq.projector}
    P(W_v(z, x),b):=\frac{1}{d} \sum_{y \in \Fd} \bar\chi(y b) W_v(yz,yx),
\end{equation}
derived in Appendix~\ref{Sec.Proofs Measurements}. Its action on vertex $v$ of a graph state can be written (up to normalization) in the form of
\begin{equation} \label{Eq.proj action}
    P(W_v(z, x),b)\ket{G} =  \ket{(z, x), b}_v U_{(z, x),b}^{(v)}\ket{G'},
\end{equation}
where the correction operator $U_{(z, x),b}^{(v)}$ is a Clifford unitary that depends on $G$, $W(z, x)$, and $b$, the eigenstate of $W_v(z,x)$ with eigenvalue $\chi(b)$ is denoted by $\ket{(z, x), b}_v$, and $G'$ is a manipulated graph without vertex $v$. In the following, we present the correction operations and resulting graphs of different local Weyl operators, and in Fig.~\ref{fig:Weyl_meas} some of these resulting graphs are exemplified. 

Since projectors do not map graph states to graph states, we introduce  
\begin{equation}
\mathcal{L} (W_v(z,x),b):=\bra{(z, x),b}_v U_{(z, x),b}^{(v)^\dag} P(W_v(z, x),b),
\end{equation}
as the measurement operation of $W_v(z,x)$ with measurement outcome $b$, such that $\mathcal{L} (W_v(z,x),b)\ket{G}=\ket{G'}$. 

\subsection{\texorpdfstring{$Z$}{Z} measurement} 
The projector associated to $Z=W(1,0)$ acting on vertex $v$ of graph state $\ket{G}$ acts according to Eq.~\eqref{Eq.proj action} with the following correction operator and manipulated graph
\begin{equation} \label{Eq.Zgraph}
    \begin{aligned}
        U_{(1,0), b}^{(v)}&=Z( b A_v), \\
        G' &= G  \setminus v.
    \end{aligned}
\end{equation}
The graph manipulation corresponds to the removal of the measured vertex and its corresponding edges.

\subsection{\texorpdfstring{$Y$}{Y}-type measurement}
Weyl operators of the form $W(1,m)$ are called $Y$ type. Their projection is Clifford equivalent to the $Z$ projection
\begin{equation}\label{Eq.Yequ}
 P(W_v(1, m),b)=L_v(\sm m)P(Z_v,b)L_v(m).
\end{equation}
The action of this projector on a graph state is given by Eq.~\eqref{Eq.proj action} with 
\begin{equation} \label{Eq.Ygraph}
    \begin{aligned}
        U_{(1,m), b}^{(v)}&= S(mA_v^2) Z(b m A'_v), \\
        G' &=  \tau_v(m)(G) \setminus v.
    \end{aligned}
\end{equation}
The resulting graph is obtained by performing a local complementation with factor $m$ on the measuring vertex, followed by the removal of said vertex and its edges.

\subsection{\texorpdfstring{$X$}{X} measurement}
The projector of the $X=W(0,1)$ measurement is local Clifford equivalent to the $W(1,1)$ projection 
\begin{equation} \label{Eq.Xequ}
    P(X_v,b)=L_{w_0}(\sm r)P(W_v(1,1),b) L_{w_0}(r),
\end{equation}
where $w_0$ is a neighbor of $v$ and $r=\sm A_{w_0,v}^{\sm 2}$. From Eq.~\eqref{Eq.Ygraph} and Eq.~\eqref{Eq.Xequ}, we obtain that the action of the projector on a graph state is encoded by 
\begin{equation} \label{Eq.Xgraph}
    \begin{aligned}
        U_{(0,1), b}^{(v)}&= R_{w_0}(\sm r) S(r\tilde{A}_{w_0}^2)  S(A'^2_v)Z( bA'_v), \\
        G' &=  \tau_v(1)(\tau_{w_0}(r)( G ))\setminus v,
    \end{aligned}
\end{equation}
where $A'$ is the adjacency matrix of $\tau_{w_0}(r)(G)$ and $\tilde{A}_{w_0}^2=A_{w_0}^2 \sm \bm e_v A^2_{w_0,v}$. This measurement, at the level of the graph, corresponds to performing a local complementation by factor $r$ on a neighboring qudit $w_0$, followed by a local complementation with factor $1$ on the measuring qudit and, subsequently, the measuring vertex is removed. Note that the choice of the neighbor $w_0$ is arbitrary, such that for a different neighbor choice, say $w_1$, one obtains a local Clifford equivalent graph state (derived in Appendix~\ref{Sec.Proofs Measurements}). 

\subsection{\texorpdfstring{$X(m)$}{X(m)} measurement}
The projector associated to $X(m)=W(0,m)$ for $m \neq 0$ is Clifford equivalent to the $X$ projector
\begin{equation}\label{Eq.Xmequ}
   P(X_v(m),b)= M_v(m) P(X_v,b)  M_v(m^{\sm 1}),
\end{equation}
such that the action on this measurement on vertex $v$ of graph state $\ket{G}$ is described by Eq.~\eqref{Eq.proj action} and
\begin{equation}\label{Eq.Xmgraph}
    \begin{aligned}
        U_{(0,m), b}^{(v)}&= R_{w_0}(\sm r) S(r\tilde{A}_{w_0}^2)  S(mA'^2_v)Z( b m A'_v), \\
        G' &=  \tau_v(1)(\tau_{w_0}(r)( G \circ_{m}v ))\setminus v,
    \end{aligned}
\end{equation}
where $w_0$ is a neighbor of $v$, $A'$ is the adjacency matrix of $\tau_{w_0}(q)( G )$, and $\tilde{A}_{w_0}^2=A_{w_0}^2 \sm \bm e_v A^2_{w_0,v}$ The action on the graph is a local multiplication by factor $m$ on the measured vertex followed by the procedure given in the $X$ measurement. 

\subsection{\texorpdfstring{$W(n,m)$}{W(n,m)} measurement}
The projector associated to $W(n,m)$ for $n \neq 0$ is Clifford equivalent to the $Z$ projector. The Clifford equivalence is obtained in two steps. First, notice that the $Z(n)=W(n,0)$ projector is equivalent to the $Z$ one via 
\begin{equation}\label{Eq.Znequ}
P(Z_v(n),b)=M_v(n^{\sm 1}) P(Z_v,b) M_v(n).
\end{equation}
Then we see that the $Z(n)$ projector is equivalent to the $W(n,m)$ projector through
\begin{equation} \label{Eq.Wnmequ}
    P(W_v(n,m), b)=L_v\left(\sm \frac{m}{n}\right)  P(Z_v(n),b)  L_v \left(\frac{m}{n}\right).
\end{equation}
The action of this projector on a graph state $\ket{G}$ yields 
\begin{equation} \label{Eq:xmznequ}
    \begin{aligned}
        U_{(n,m), b}^{(v)}&=S\left(\frac{m A^2_v}{n}\right)Z( b A'_v), \\
        G' &=  \tau_v\left(\frac{m}{n}\right)( G )\circ_{n^{\sm 1}}v \setminus v,
    \end{aligned}
\end{equation}
where $A'$ is the adjacency matrix of the graph $\tau_v(m/n)( G )$. From Eq.~\eqref{Eq:xmznequ}, one can identify that the graph gets transformed via first performing a local complementation by factor $m/n$ on the measuring vertex, followed by a local multiplication with factor $n^{\sm 1}$ on said vertex and last, the removal of the measured vertex.

\section{Qudit Noisy Stabilizer Formalism} \label{Sec.NSF}
In this section, we introduce the qudit noisy stabilizer formalism (NSF$_d$), a method to efficiently describe and track Pauli-diagonal noise acting on a graph state. The standard density matrix description is prohibitively costly, due to the necessity of working with matrices that grow exponentially with the number of qudits, i.e., $d^n\times d^n$. As previously described in Sec.~\ref{Subsec.Noisy stabilizer formalism}, the qubit instance of this formalism treats the noise channels and the graph state individually, avoiding operations involving exponentially big matrices. This approach is also followed here for the qudit case and is also equivalent to the density-matrix-based one. 

Consider a graph state $\rho=\ketbra{G}$ subject to Pauli-diagonal noise channels $\mathcal{E}_1 \cdots \mathcal{E}_m$, where we apply a series of manipulation operators $O_l \cdots O_1$, which are Clifford or Weyl measurement operators (as the ones presented in Sec.~\ref{Sec.Manipulation of graph states} and Sec.~\ref{Sec.Weyl measurements of graph states}), such that 
\begin{equation}\label{eq:noisy:graph:state}
    \left(O_l\cdots O_1\right)\mathcal{E}_1 \cdots \mathcal{E}_m\rho \left(O_l\cdots O_1\right)^{\dagger}.
\end{equation} 
With this formalism, we update the noise channels and the pure graph state for each manipulation operation via an efficient set of rules. Such that we end with a series of updated noise channels that act on the noiseless manipulated graph state. 

\subsection{Update rules} \label{Sec.Update rules} 
Here we present the core of the NSF$_d$, namely the update rules for each individual noise operator. Consider a Clifford or Weyl measurement operator $O$, which would correspond to $O_1$, and a generalized Pauli noise operator term $N$ from a channel $\mathcal{E}_j$ from the channel sequence $\mathcal{E}_1 \cdots \mathcal{E}_m$ such that the manipulation is applied on a noisy graph state, $ON\ket{G}$. The two operators commute up to a change of the noise operator, and we obtain $\tilde{N}O\ket{G}$; this change is given by the update rules.

As previously described in Sec.~\ref{Subsec.Noise model}, any arbitrary generalized Pauli noise operator acting on a graph state $\ket{G}$ can be translated into a product of $Z$-type operators. Therefore, it suffices to study only the update rules of $Z$-type noise operators. Furthermore, phases appearing in the computation of the update rules are negligible, since we consider Pauli-diagonal channels, where such phases cancel. Below, we showcase the building-block update rules needed to generate the update rule of any manipulation operators via the concatenation of said building-block update rules. In Fig.~\ref{fig:update_rule}, an illustration of such a concatenated update rule is shown. See Appendix~\ref{Sec.Proofs:Update rules} for the derivation of the given update rules.

\subsubsection{Clifford operators}
The Clifford group acting on graph states is generated by the local complementation, the local multiplication, and the $CZ$ gate. All these operations map graph states to graph states.

\begin{figure}
    \centering
    \includegraphics[width=\columnwidth]{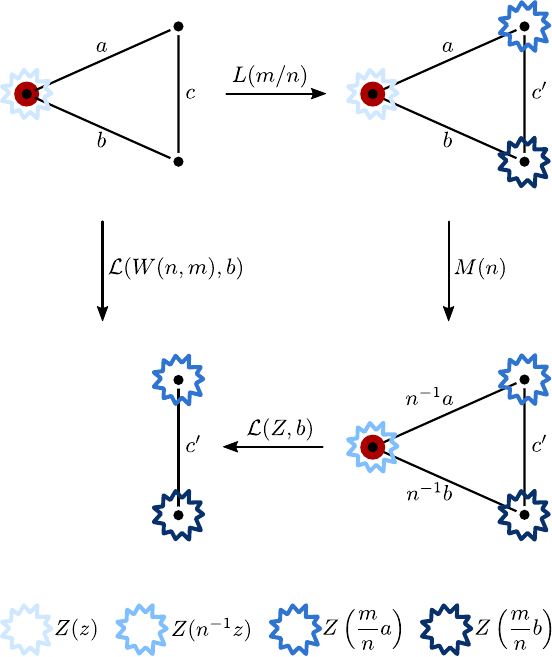}
    \caption{Graphical representation of the $W(n,m)$ measurement update rule constructed by concatenating the building-block update rules, which are applied in the following order: first $L(m/n)$, followed by $M(n)$, and last, a $Z$ measurement. The vertex where the operations are applied is shaded dark red. The jagged lines depict $Z$-type noise operators. The weight $c'$ is $c+mn^{\sm 1}ab$.}
    \label{fig:update_rule}
\end{figure}

The local multiplication $M(m^{\sm 1})$ and the $Z(z)$ Weyl operators commute if they act on different vertices. However, if they act on the same vertex, then the update rule is
\begin{equation} \label{Eq.ur_local_mult}
    M_v(m^{ \sm 1}) Z_v(z) \ket{G}= Z_v(m z)  M_v(m^{ \sm 1})\ket{G}.
\end{equation}
The local complementation $L(m)$ and the $Z(z)$ Weyl operators commute when acting on different vertices, but when acting on the same vertex, they exhibit the following update rule
\begin{equation}\label{EqLCupdate}
     L_v(m) Z_v(z)  \ket{G} = Z_v(z) Z(A_v mz)  L_v(m) \ket{G}.
\end{equation}
These two update rules are depicted in Fig.~\ref{fig:update_rule}. Last, the $CZ$ gate commutes with any $Z(z)$ operator, so there is only the trivial update rule. 

\subsubsection{Weyl measurements}
As seen in Sec.~\ref{Sec.Weyl measurements of graph states} the measurement operations are all equivalent up to Clifford operations. Therefore, it suffices to have the update rule for one measurement operation, here the Weyl $Z$ measurement such that the update rule for the rest of the Weyl measurements can be derived by concatenating the update rules of the corresponding Clifford operations derived above and of the Weyl $Z$ measurement, as presented and derived in Appendix~\ref{Sec.Proofs:Update rules} and exemplified in Fig.~\ref{fig:update_rule}.

The Weyl $Z$ measurement $\mathcal{L} (Z_v,b)$ and the $Z_v$ operator have the update rule
\begin{equation} \label{Eq.ur_z_meas}
    \mathcal{L} (Z_v,b) Z_v(z) \ket{G}= \mathcal{L} (Z_v,b)\ket{G},
\end{equation}
and if they act on different vertices, then they commute. 

Importantly, one can see from Eq.~\eqref{EqLCupdate} that the noise operator acting on the manipulated qudit spreads throughout the graph. Therefore, sequential manipulations that include local complementations, e.g., the $Y$-type measurement and $X$ measurement, that act on neighboring qudits on a noisy graph state do not commute. This implies that different orders in which certain manipulations are performed lead to different final noise patterns if the manipulations are performed on neighboring qudits, as also observed in the qubit case in Ref.~\cite{1-2023}.

\subsection{Methodology}\label{ssec:method}
Given the initial noisy state described by Eq.~\eqref{eq:noisy:graph:state}, with the NSF$_d$ we efficiently describe and track the noise channels and the graph state.

The considered noise channels are Pauli-diagonal, so they commute with each other, and thus, without loss of generality, we can write the noise channels as $Z$ type, as they act on a graph state (see Sec.~\ref{Subsec.Noise model}). This concatenation of commuting noise channels also commutes with a manipulation operator, up to a change of the noise terms of the channel, reflected in the update rules. Hence, we can update the noise channels $\mathcal{E}_i$ individually. The method is such that the update is done for each manipulation operator iteratively. 
\begin{itemize}
\item[1.] So one starts with computing the effect of $O_1$ on both the noise channels and the noiseless graph state, leading to $\tilde{\mathcal{E}}_i$ and $\rho'=\ketbra{G'}$, respectively. For each channel, for each noise term, each Pauli $Z$ contribution is updated using the above update rules for $O_1$. The update of the graph state is efficiently tractable thanks to the generalized Gottesman-Knill theorem \cite{2-2013}. The state after updating for $O_1$ is 
\begin{equation*}
    (O_l \cdots O_2) \tilde{\mathcal{E}}_1 \cdots \tilde{\mathcal{E}}_m \rho' (O_l \cdots O_2)^{\dagger}.
\end{equation*}
\item[2.] Then the same procedure is performed for the next manipulation operator $O_2$, where now one updates the noise channels $\tilde{\mathcal{E}}_i$ and the noiseless graph state $\rho'=\ketbra{G'}$, leading to $\tilde{\tilde{\mathcal{E}}}_i$ and $\rho''=\ketbra{G''}$. 
\begin{equation*}
    \vdots
\end{equation*}
\item[$l$.] This is performed for the rest of the manipulation operators to retrieve the final updated noise channels, denoted as $\widetilde{\mathcal{E}}_i$, and graph state, so that the noisy graph state is
\begin{equation*}
    \widetilde{\mathcal{E}}_1 \cdots \widetilde{\mathcal{E}}_m \left(O_l\cdots O_1\right) \rho \left(O_l\cdots O_1\right)^{\dagger},
\end{equation*}
and is equivalent to the state in Eq.~\eqref{eq:noisy:graph:state}. 
\end{itemize}

Note that the used update rules depend at each step on the current graph state, where the noise is acting on. Therefore, one individually updates the channels and the state, but the update of the channel depends on the graph state. 

Moreover, when a manipulation operator corresponds to a measurement, both the size of the graph state and the size of the noise channels can be reduced by one vertex. Furthermore, as mentioned above, the order of Weyl measurements on consecutive qudits is of relevance, as opposed to the noiseless case, where Weyl measurements on different qudits commute. However, this does not imply that measurements cannot be performed at the same time. Consider a certain sequence of Weyl measurements acting on neighboring qudits $\mathcal{L} (W_{v_l}(z_l,x_l),b_l) \cdots \mathcal{L}(W_{v_1}(z_1,x_1),b_1)$, one can find a \textit{measurement pattern} that can be applied simultaneously by commuting all correction operators of the Weyl measurements to the last step. So we define $\mathcal{U}=U_{(z_l, x_l),b_l}^{(v_l)^\dag}\cdots U_{(z_1, x_1),b_1}^{(v_1)^\dag}$ and $\mathcal{P}=\widetilde{P}(W_{v_l}(z_l, x_l),b_l)\cdots \widetilde{P}(W_{v_1}(z_1, x_1),b_1)$ such that the sequence of measurements corresponds to $\mathcal{U}\mathcal{P}$ and $\mathcal{P}$ denotes a set of Weyl projectors that commute with each other and can be applied in any order and is denoted as measurement pattern. This pattern corresponds to a certain order in the NSF$_d$ description, and this order is encoded in $\mathcal{U}$. The translation between the order given by the NSF$_d$ and a measurement pattern has been worked out for the qubit case in Ref.~\cite{MorRuiz2025imperfectquantum}. Note that updating the noise channel for this given sequence of measurements can be efficiently done as described above.

\section{Efficiency} \label{Sec.Efficiency}
Given an $n$-qudit graph state subject to a general Pauli-diagonal channel, as the one described in Eq.~\eqref{eq:general:channel}, one has to compute the update of all $Z$-type operators for each qudit, since as discussed in Sec.~\ref{ssec:method} all noise terms of all noise channels can be treated individually. Therefore, the number of different update rules one has to compute is at most $d\times n$ when all noise operators are translated to the $Z$ type, thanks to the stabilizer structure of graph states. 

Besides the computational complexity of the update rules, one also has to account for the number of terms that have to be updated. Given that all noise maps and noise operators are treated individually, one has to update as many noise operators as the sum of the number of noise terms of the individual channels. This is in contrast to the case where noise channels could not be treated individually, i.e., one would multiply the channels out, which in general would lead to exponentially many terms. For example, given $n$ single-qudit noise maps acting on different qudits, one needs to consider $d^2 \times n$ noise terms with the independent treatment of noise channels, whereas if one multiplies them out, then the number of terms is $d^{2n}$.

The main bottleneck of the NSF$_d$ is the size of the final updated state, since to apply the final updated noise maps to the corresponding noiseless state, one has to operate with density matrices of the size of $d^m\times d^m$, where $m$ denotes the final size.

\section{Generalizations} \label{Sec.Generalizations}
In Sec.~\ref{Sec.NSF}, the NSF$_d$ for qudit graph states of prime-power dimension subject to Pauli-diagonal noise is presented. In the following, we investigate extensions of the method and examine its limitations by applying it to broader quantum systems and noise models.

\subsection{Stabilizer states}
The presented NSF$_d$ can be extended to the stabilizer states, as prime-power dimensional stabilizer states are local Clifford equivalent to graph states such that, given a stabilizer state $\ket{S}$, we can write the state as $U \ket{G}$, with $U$ being a local Clifford and $\ket{G}$ being a graph state. Consider that the stabilizer state is subject to a noise operator $N$, then we use the commutation relations from the local Clifford and the noise operator to obtain $N \ket{S}=N U \ket{G}= U \tilde{N} \ket{G}$. 

Since Clifford operations map stabilizer states to stabilizer states, the action of a Clifford $C$ on a stabilizer state yields $C\ket{S} = U' \ket{G'}$, where $U'$ is a local Clifford and $G'$ is a graph. Therefore, a noisy stabilizer state transforms under the action of a Clifford $C$ as $CN\ket{S}=\tilde{N}' U' \ket{G'}$, where we use the commutation relation of the noise operator and the Clifford $C$ such that it leads to the updated noise operator $\tilde{N}'$.

When performing Weyl measurements on stabilizer states, we can restrict to $Z$ measurements since other Weyl measurements are local Clifford equivalent to it. A noisy stabilizer state transforms under the action of a $Z$ measurement as $P(Z,b)N\ket{S}= P(Z,b) U \tilde{N} \ket{G}=U P(W,\tilde{b}) \tilde{N} \ket{G}$, where we have used that $U$ and $P(Z,b)$ commute up to a change of the Weyl basis of the measurement. Finally, this noisy state can be described and updated with the graph state NSF$_d$ from above.

Last, for both general Clifford manipulations and Weyl measurements, we use the described commutations and the previously described methodology.

Alternatively, one can directly work with stabilizer states without explicitly knowing their locally Clifford equivalent graph states. Under Clifford operations, noise channels can be updated by conjugating each noise operator individually, independently of the state.

For Weyl measurements, we require the error channels to commute with the Weyl projector, which can be ensured by locally adjusting the noise operators via inserting stabilizer operators. Specifically, if for the measurement vertex $v$, then we have stabilizer operators $S^{i} \in \mathcal{S}$ such that $\{ S_v^{i} \}=\{ W_v(z,x)| z,x \in \Fd \}$ (up to a phase), these adjustments are straightforward. Such stabilizers always exist unless vertex $v$ is disconnected in the local Clifford equivalent graph. Indeed, for a connected vertex $v$ of a graph state, each local Weyl operator can be generated by multiplying suitable stabilizer generators from vertex $v$ with those of a neighboring vertex. Since local Clifford operations map the local Weyl group onto itself, the same property extends to the local Clifford equivalent stabilizer state. Thus, the absence of stabilizers generating the local Weyl operators implies disconnection of the vertex, in which case noise operators on the measurement vertex can simply be locally removed after the measurement, as dictated by the update rules derived from graph states. 

This approach does not yield a fixed set of update rules, as is the case for the graph-state-based method; instead, noise updates are computed dynamically.

\subsection{Arbitrary finite dimension} \label{SubSec.Arbitrary dimension}
Throughout this work, we have restricted ourselves to systems of local prime-power dimensions. This is due to the fact that composite dimensional stabilizer states are, in general, not Clifford equivalent to graph states \cite{GHEORGHIU2014505}. Therefore, the reduction of the stabilizer case to the graph state case presented above fails. Nonetheless, using the linearized stabilizer formalism introduced in \cite{1-2023}, one can obtain a form of a \textit{qudit noisy graph state formalism} analog to the one in the prime-power case for composite dimensions. For the construction of said analog, we defer to Appendix~\ref{Sec.Qudit graph state formalism for arbitrary finite dimensions}.

\subsection{General channels} \label{SubSec.General channels}
The NSF$_d$ can be applied when dealing with any Pauli-diagonal noise channel. Channels including off-diagonal Pauli terms do, in general, not commute. Therefore, one cannot simply translate the noise channels to $Z$-type channels, since one would need to multiply out the error channels in order to be able to act on the graph state directly. For a noise channel $\mathcal{E}$ with not all $Z$-type operators, its commutation with the $Z$ projector leads to a possible different change of the projector for each noise term such that $P(Z,b)\mathcal{E} \ketbra{G} \neq \tilde{\mathcal{E}}P(W,\tilde{b}) \ketbra{G}$. So, to treat non-Pauli-diagonal noise channels, one cannot treat the noises independently, i.e., one has to multiply out the terms of the noise channels, and the computation is no longer efficient. 

Nevertheless, as previously mentioned, one can enforce a Pauli-diagonal form via the random application of unitary gates before and after the application of the noise channel, without changing the diagonal elements or the Choi-Jamio\l kowski fidelity of the noise channel \cite{dur2005standard}.

\section{Application of method}\label{Sec.Application of method}
In this section, we illustrate how the noisy stabilizer formalism can be applied to an example and demonstrate the power of the formalism.
\subsection{Setting}
\begin{figure}[h]
    \centering
    \includegraphics[width=1\linewidth]{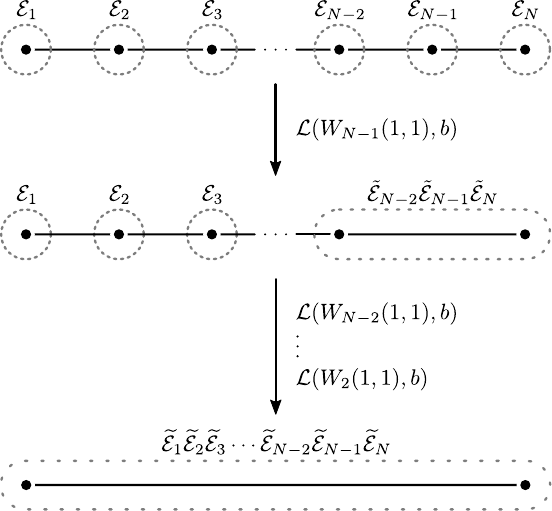}
    \caption{Depiction of qudit linear chain with local depolarizing noise. The edge weights are constant one and therefore omitted. The middle qudits are measured out via $W(1,1)$ measurements, obtaining at the end a generalized noisy Bell pair.}
    \label{fig:application}
\end{figure}

As the initial graph state, we consider an $N$-qudit open linear cluster with edges of weight one. The associated graph $G=(V,E,A)$ has vertices labeled from 1 to $N$, such that $V=(1,2,\dots,N)$, and the elements of the adjacency matrix are defined by $A_{ij}=\delta_{1,|i \sm j|}$. The aim is to obtain a generalized Bell pair between the ends of the cluster, $1$ and $N$, by performing $W(1,1)$ measurements on the middle qudits, $(2, \dots, N \sm 1)$, as depicted in Fig.~\ref{fig:application}. 

We consider that the qudits in the linear cluster are subject to two sources of noise, one where the noise probability is independent of the dimension of the system, and the other is dimension dependent, these dependencies of error rates are present in the experimental realization of high-dimensional systems on trapped ions \cite{ringbauer2022universal}. We model both sources as single-qudit depolarizing noise, whose action on qudit $v$ of a noiseless graph state $\rho$ is
\begin{equation}\label{Eq.Depol}
\mathcal{E}_v(\lambda) \rho = \lambda \rho + \frac{1 \sm \lambda}{d^2} \sum_{z,x \in \Fd} Z_v(z) Z(\sm x A_v)\rho Z(x A_v) Z_v(\sm z). 
\end{equation}

The dimension-independent noise model is modeled by $\mathcal{E}_v(\lambda=r)$. The dimension-dependent one is modeled by $\mathcal{E}_v(\lambda=q_d)$, where the dimension dependency is encoded in the parameter $q_d$. We consider systems of dimension $d=2^m$ and we determine said dependency by matching the Choi-Jamio\l kowski fidelity (see Sec.~\ref{sec:Choi-Jamiolkowski}) of $m$ qubit depolarizing channels, each written as $\mathcal{E}_v(\lambda=q_2)$, to the Choi-Jamio\l kowski fidelity of one $2^m$-dimensional depolarizing channel, written as $\mathcal{E}_v(\lambda=q_d)$. We note that the Choi-Jamio\l kowski fidelity of the depolarizing channel\footnote{This corresponds to the probability of identity acting on the system.} in Eq.~\eqref{Eq.Depol} is given by $\lambda+(1 \sm \lambda)/d^2$, and the Choi-Jamio\l kowski fidelity of multiple channels is just the multiplication of individual ones. Hence, via equating the Choi-Jamio\l kowski fidelities of a $2^m$-dimensional depolarizing noise channel and $m$ qubit depolarizing noise channels, we obtain
\begin{equation} \label{Equ:tildef}
    q_{d}=\frac{(3q_2+1)^m \sm 1}{4^m \sm 1},
\end{equation}
as the adapted depolarizing parameter for the dimension-dependent depolarizing channel $\mathcal{E}_v(\lambda=q_d)$. In Appendix~\ref{Sec.Physically motivated noise models} we motivate this choice further and also discuss different dimensional dependencies of the depolarizing parameter.

The concatenation of the two depolarizing channels $\mathcal{E}_v(\lambda=r)$ and $\mathcal{E}_v(\lambda=q_d)$ corresponds to the depolarizing channel $\mathcal{E}_v(\lambda=r \times q_d)$ such that the initial noisy linear cluster is $\mathcal{E}_1(\lambda=r \times q_d)\cdots \mathcal{E}_N(\lambda=r \times q_d) \rho$.

\subsection{Analysis} \label{Sec.Analysis}
The manipulation of a linear cluster subject to single-qudit depolarizing noise, modeled by Eq.~\eqref{Eq.Depol} with $\lambda=r\times q_r$, into a generalized Bell pair between the ends can be done by performing $W(1,1)$ measurements on the middle qudits. Note that as mentioned in Sec.~\ref{Sec.NSF}, the order of consecutive measurements leads to different final noise maps. For an arbitrary order of said measurements, using the update rules one obtains the updated noise channels, denoted by $\widetilde{\mathcal{E}}_i$\footnote{For simplicity we do not write the argument with the parameter $\lambda$, as for this case it is fixed to $\lambda=r\times q_r$.}, which act on the noiseless generalized Bell pair, denoted by $\rho'$. 

The updated noise maps of the edge qudits, $1$ and $N$, are
\begin{equation}\label{eq.mtarget}
\begin{aligned}
&\widetilde{\mathcal{E}}_1\rho'=\widetilde{\mathcal{E}}_N\rho'= \\
&\lambda \rho'+\frac{1 \sm \lambda}{d^2} \sum_{z,x \in \Fd}
Z_1(z) Z_N(x) \rho' (Z_1(z) Z_N(x))^{\dag}.
\end{aligned}
\end{equation}
We find that the updated noise maps of the middle qudits, $(2, \dots , N \sm 1)$, are of the form 
\begin{equation} \label{eq.minner}
\begin{aligned}
&\mathcal{M}(\alpha,\beta)\rho'=\\
&\lambda \rho' +\frac{1 \sm \lambda}{d}\sum_{u \in \Fd} Z_1(\alpha u) Z_N(\beta u) \rho' (Z_1(\alpha u) Z_N(\beta u))^{\dag},
\end{aligned}
\end{equation}
where $\alpha, \beta \in \Fp$. The corresponding derivations are presented in Appendix~\ref{SubSec.Form of final noise maps}. 

Notice that applying $w$ times a map of the shape of Eq.~\eqref{eq.minner} with parameter $\lambda$ is the same as applying it once with parameter $\lambda^w$. So we define the weight of a map $\mathcal{M}(\alpha,\beta)$ as how many times it is applied to the final state, and we denote it as $w_{\alpha}^{\beta}$. We can collect all weights into a so-called \textit{weight vector} as follows
\begin{equation}
\pmb{w}:=[w_0^0,w_0^1,\dots,w_0^{p\sm 1},w_1^0,w_1^1,\dots,w_1^{p\sm 1},\dots,w_{p \sm 1}^{p \sm 1}].
\end{equation}
The elements of the weight vector depend on the chosen measurement order. A simple choice is the strategy of measuring the middle qudits sequentially from $N\sm 1$ to $2$, i.e., right to left. This side-to-side strategy yields the updated noise maps of the form $\widetilde{\mathcal{E}}_j=\mathcal{M}(1, j \sm 1)$ for $j\in\{2, \dots, N \sm 1\}$, which can be described by the following weight vector
\begin{equation}\label{eq:sidetoside}
\pmb w = [\bm 0_p, \bm m_p+\sum^{s}_{j=1} \bm e_j,\bm 0_{p^2 \sm 2p}],
\end{equation} 
with $n:= N\sm 2$, $n=mp+s$, $m,s \in \mathbb{N}_0$, $s<p$, and $\bm x_n$ refers here to a vector of size $p$ with the value $x$ in each entry, except $\bm e_j$ which refers to the standard unit vector. The derivation of this strategy is given in Appendix~\ref{SubSec.Side to side strategy}.

With the updated final noise maps, we can compute the fidelity of the final generalized Bell pair. In Appendix~\ref{SubSec.Fidelity of final state} we compute an analytic formula of the fidelity of the generalized Bell pair for an arbitrary measurement strategy applied on an $N$-qudit linear cluster with depolarizing probability $1 \sm \lambda$, which leads to a weight vector $\bm w$. In order to do this calculation, we compose all updated noise maps such that only the noise terms proportional to the identity of this composite noise map contribute with a nonzero value to the fidelity. Thus, this formula depends on the weight vector $\bm w$, the depolarizing probability $1 \sm \lambda$, and the dimension $d$. 

\subsection{Results} \label{Sec.Comparison of different dimensions}
Our goal is to investigate how the fidelity of the final generalized Bell state obtained from the manipulation using the side-to-side strategy of a linear cluster subject to two noise sources, one dependent on the local dimension and the other not, is affected by the local dimension. To this aim, we compare one system of local dimension $2^m$ with $m$ systems of local dimension $2$. In order to facilitate a fair comparison of the fidelities of different dimensional systems, we raise the qudit fidelity of dimension $2^m$ to the power of $1/m$ such that the adapted fidelities of different dimensional generalized Bell pairs mean the same in terms of entanglement. 

\begin{figure}[h]
    \centering
    \includegraphics[width=1\linewidth]{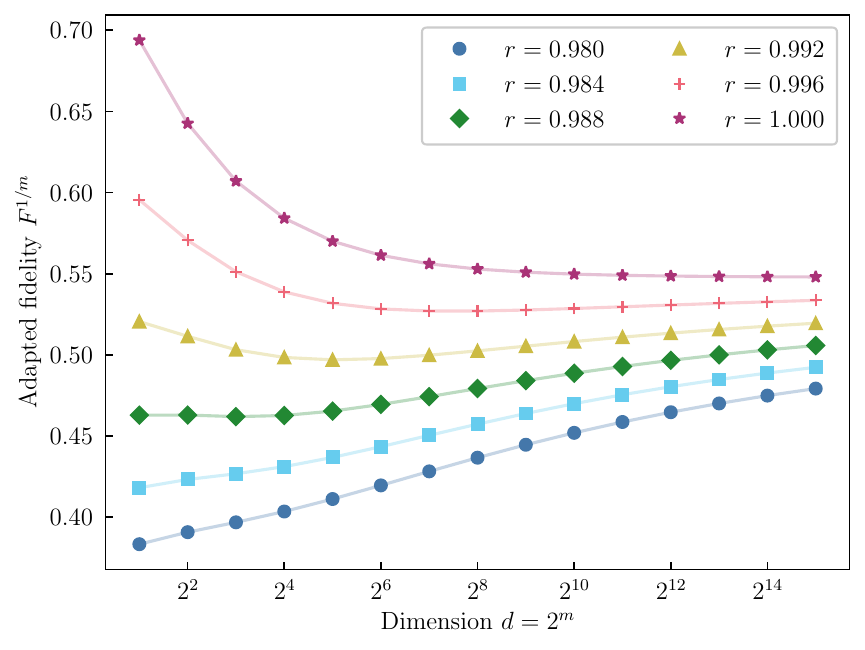}
    \caption{Adapted fidelity $F^{1/m}$ of a generalized Bell pair obtained via the side-to-side strategy from a noisy linear cluster of size $N=100$ in terms of the local dimension $d=2^m$, for different dimension-independent depolarizing parameters $1 \sm r$. The dimension-dependent depolarizing noise with strength $1-q_d$ scales with the dimension as $q_d=((3q_2+1)^m \sm 1)/(4^m \sm 1)$, with $q_2=0.992$.}\label{fig:fidelity_choi}
\end{figure}

Figure~\ref{fig:fidelity_choi} illustrates the behavior of the adapted fidelity as a function of the local dimension for different values of the dimension-independent parameter governing the strength of the depolarizing channel, $1 \sm r$. When the dimension-independent depolarizing effect is strong, i.e., low $r$, the adapted fidelity exhibits an increasing trend with higher local dimensions. Conversely, for a weaker dimension-independent depolarizing influence, i.e., high $r$, the adapted fidelity decreases as the dimension grows. In an intermediate regime, the adapted fidelity initially declines before exhibiting a subsequent increase with increasing dimension. These effects are due to the interplay of the two considered noise sources, as one becomes more dominant with increasing dimension and the other one stays constant. One can see that the final generalized Bell pair has the highest adapted fidelity for either $d=2$ or $d \rightarrow \infty$. However, depending on the studied dimension dependency of the depolarizing parameter, one can find an optimal intermediate dimension. In Appendix~\ref{Sec.Physically motivated noise models} we present different dimensional dependencies that show optimal intermediate dimensions.

\section{Conclusions and outlook} \label{Sec.Conclusions and outlook}
In this paper, we generalize the noisy stabilizer formalism to systems of local prime-power dimension such that one can describe qudit systems subject to Pauli-diagonal noise and update them under Clifford operations and local Weyl measurements efficiently. The key concept lies in the separate treatment of noise channels and pure states, avoiding the necessity of working with density matrices. This separation is enabled by defining rules for updating error channels and efficiently updating stabilizer states under Clifford operations and generalized Pauli measurements. To derive these rules, we utilize prime-power dimensional graph states, where we additionally derive a set of graphical rules for local Weyl measurements on such graph states. Finally, we demonstrate the utility of the proposed formalism by analyzing the fidelity of a generalized Bell pair generated from a noisy linear chain of qudits.

Given the prevalence of quantum protocols that utilize qudit graph states under the influence of Clifford operations and Weyl measurements, and considering the role of Pauli-diagonal noise as a general noise model, the qudit noisy stabilizer formalism offers a systematic framework for analyzing the impact of noise on these protocols. Protocols employing qudit graph states range from quantum error correction \cite{PhysRevA.84.052306,PhysRevA.105.032450} to entanglement purification \cite{colrecurrence} to measurement-based quantum computation of Clifford circuits \cite{PhysRevA.68.062303,PhysRevA.95.032312}. 

In addition to the qudit noisy stabilizer formalism for prime-power dimensions, we also introduce the qudit noisy graph state formalism for arbitrary dimensions. This formalism applies specifically to graph states transforming under Weyl measurements and Clifford operations that map graph states to graph states. Since the stabilizer formalism exists not only for finite dimensional systems, but also for continuous variable systems \cite{RevModPhys.77.513}, with a continuous variable version of the Gottesman-Knill theorem \cite{PhysRevLett.89.207903} and a stabilizer-state graph-state equivalence \cite{PhysRevA.80.052333}, the formulation of a continuous variable noisy stabilizer formalism is feasible and we will report it on a separate work \cite{Aigner25}.

Nevertheless, the qudit noisy stabilizer formalism is limited (in its full generality) to prime-power dimensions and to noise that is diagonal in the Heisenberg-Weyl basis. Generalization to nondiagonal noise processes seems generally infeasible, due to the noncommutativity of the error channels, which leads to the impossibility of independent treatment of the channels, which is key for the efficiency of the formalism. Last, obtaining the final state at the end of an operation sequence requires working with density matrices of the size of the final state, which imposes an exponential overhead on the size of the final state on the formalism, even though the individual updates of the state and error channels are efficient. However, if the final state is small, i.e., consists only of a few qudits, then the qudit noisy stabilizer formalism offers a robust framework for analyzing the impact of noise and investigating quantum information processing protocols in the presence of imperfections.

\section*{Acknowledgments}
The authors acknowledge support by the Austrian Research Promotion Agency (FFG) under Contract No. FO999914030 (Next Generation EU) and also that research was funded in whole or in part by the Austrian Science Fund (FWF) 10.55776/P36009, 10.55776/P36010, and 10.55776/COE1. For open access purposes, the authors have applied a CC BY public copyright license to any author-accepted manuscript version arising from this submission. Finanziert von der Europäischen Union.

\normalem 

\bibliographystyle{apsrev4-1}
\interlinepenalty=10000
\bibliography{refs.bib}
\interlinepenalty=10000
\clearpage

\renewcommand\appendixname{Appendix}
\appendix
\onecolumngrid

\section{Background for even prime-power dimension}
\label{Sec. Background for even prime power dimension}
In Sec.~\ref{Sec.Background}, we have introduced the necessary mathematical background for odd prime-power dimensions. Here we present the differences with the even prime-power dimensions for the generalized Pauli group, the stabilizer formalism, and the Clifford group \cite{1-2021}. Notice that graph states for even dimensions can be defined analogously to the odd case \cite{grassl2002graphs,schlingemann2001stabilizercodesrealizedgraph}.

\subsection{Generalized Pauli group}
The generalized Pauli $X$ operator is the same as for the odd case, given by Eq.~\eqref{eq:pauli:x}. The $Z$ operator is defined as
\begin{equation}
    Z(z):=\sum_{y \in \Fd} \chi_4(2yz) \ketbra{y},
\end{equation}
where $\chi_4(x)=i^{\text{tr}(x)}$ and $\text{tr}(t)$ is the trace of the linear map on the $\mathbb{Z}_4$-module $R_{4^m}$ which acts as $x \mapsto x \cdot t$.
Similar to the odd case, the Weyl operator for even prime-power dimensions is defined as
\begin{equation}
    W(\bm z,\bm x,t):= \chi_4(t \sm \bm z \bm x) Z(\bm z)X(\bm x),
\end{equation}
and the Heisenberg-Weyl group is the group containing all these elements. Throughout this text, we use the following overloaded notation, $\chi(x):=\chi_4(2x)$, for the even case, since $\chi(x)$ and $\chi_4(2x)$ are both homomorphisms. 

\subsection{Stabilizer formalism}
Here an isotropic subspace, Eq.~\eqref{eq:isotropic} for the odd case, $K \subset \Fd^{2n}$ with lift $\alpha$ defines $K^{\alpha}:= \{(\bm v,2\alpha(\bm v))| \bm v \in K \}$,
which in turn sets the Abelian subgroup $W(K^\alpha):=\{\omega^{\text{tr}(\alpha(\bm v))} W(\bm v)| \bm v \in K \}$. To describe the lift $\alpha$, we introduce 
\begin{equation}
\beta(\bm v,\bm w)= \gamma(\bm v+ \bm w) \sm \gamma(\bm v) \sm \gamma(\bm w)+2(\bm v_z \bm w_z), \quad \gamma(\bm v):=\bm v_z  \cdot \bm v_x,
\end{equation}
where the entries of $ \bm v_z$ and $\bm v_x$ are viewed as elements of the Galois ring $\mathbbm{GR}_{4^m}$ by identifying each coefficient of the polynomial representation of finite field elements as an element of $\mathbb{Z}_4$. This is done such that the lift $\alpha$ is defined via the following relations 
\begin{equation}
\begin{aligned}
    2\bar \beta (\bm v,\bm w)&= \beta(\bm v,\bm w), \\
    \bar \beta(\bm v,\bm w)&= \alpha(\bm v+ \bm w) \sm \alpha(\bm v) \sm \alpha(\bm w), \\ 
    \alpha(\bm 0)&=0.
\end{aligned}
\end{equation} 
When choosing a basis on $K$, $\{\bm v_1,\dots, \bm v_k\}$, we pick $\alpha(\bm v_i \lambda)=0, \forall \lambda \in \Fd$ such that $\alpha$ is unique. A Lagrangian $\mathbbm{L}$ uniquely defines a stabilizer state $\ket{\mathbbm{L},\bm v}$ by the following eigenvalue equation
\begin{equation}  
    \omega^{tr(\alpha(\bm w))} \chi([\bm v,\bm w])W(\bm w)\ket{\mathbbm{L},\bm v}= \ket{\mathbbm{L},\bm v} \ \forall \bm w \in \mathbbm{L}.
\end{equation} 

\subsection{Clifford operations}
The generators of the Clifford group are the same as the ones presented in Table~\ref{Table:1}, except the $S$ operator, which is defined as 
\begin{equation} 
S:= \sum_{x\in \Fd} \chi_4(x^2)\ket{x}\bra{x}.
\end{equation} 
Moreover, if the power of our prime-power dimension $m$ is odd, then the $H$ operator is defined as $H_{\text{even}}=e^{i \pi /4} H_{\text{odd}}$. 
As in the odd case, the Clifford group normalizes the Heisenberg-Weyl group and maps stabilizer states to stabilizer states. 

\section{Proofs for Weyl measurements of graph states}
\label{Sec.Proofs Measurements}
In this section, we prove the results regarding the Weyl measurements on graph states presented in Sec.~\ref{Sec.Weyl measurements of graph states}. For the derivation, we make use of the commutation relations between the Weyl operators and the Clifford generators presented in Table~\ref{Table:commutation_relations}, and we write c.r. to denote their use.

Here we show how to obtain the expression of the single-qudit Weyl measurement Eq.~\eqref{Eq.projector} presented in the main text, from projectors defined via isotropic subspaces. In the odd prime-power case the projector associated to the isotropic subspace $K$ and vector $\bm v \in \Fd^{2n}$ is given by
\begin{equation}\label{eq:projector:odd}
    P(K,\bm v)=\frac{1}{|K|}\sum_{\bm w \in K} \chi([\bm v,\bm w])W(\bm w).
\end{equation}
In the even dimension case (see Appendix~\ref{Sec. Background for even prime power dimension}), the projector associated to the isotropic subspace $K$ vector $v$ and lift $\alpha$ is given by
\begin{equation}\label{eq:projector:even}
    P(K,\bm v,\alpha)=\frac{1}{|K|}\sum_{\bm w \in K} \chi([\bm v,\bm w]) \omega^{\text{tr}(\alpha(\bm w))} W(\bm w).
\end{equation}
In particular, if we measure a single-qudit Weyl operator $K=\langle \bm a \rangle $, then $\alpha(\bm v)=0, \bm v \in K$ and we have the same projector associated to a measurement as in the odd case 
\begin{equation}\label{Eq.projector2}
    P(\langle \bm v \rangle ,\bm w)=\frac{1}{|\langle \bm v \rangle |}\sum_{y \in \Fd} \chi(y [\bm v, \bm w])W(y \bm v),
\end{equation} 
which is the measurement of a Weyl operator $W(\bm v)$ with measurement outcome $[\bm v, \bm w]$. For a local Weyl operator $W_v(z, x)$ the projector can be written using Eq.~\eqref{Eq.projector2} as
\begin{equation}
P( \langle (\bm e_v z, \bm e_v x) \rangle,\bm w) = \frac{1}{d} \sum_{y \in \Fd} \bar\chi(y b) W(y (z \bm e_v , x \bm e_v )),
\end{equation} 
with $b=[\bm w,(z\bm e_v , x\bm e_v )]$ being the measurement outcome, with which we obtain the expression of the single-qudit Weyl measurement presented in the main text Eq.~\eqref{Eq.projector}.

\begin{table}[h]
\centering
\begin{tabular}{c|ccl|ccl}
Clifford operator $U$ &
  \multicolumn{3}{c|}{$U Z_j(z) U^{\dagger}$} &
  \multicolumn{3}{c}{$U X_j(x) U^{\dagger}$} \\ \hline\hline
\multirow{2}{*}{$H_v$}          & $Z_j(z)$              & for    & $j\neq v$    & $X_j(x)$                                       & for    & $j\neq v$   \\
                                & $X_v(\sm z)$             & for    & $j=v$        & $Z_v(x)$                                       & for    & $j=v$       \\ \hline
\multirow{2}{*}{$H_v^\dag$}     & $Z_j(z)$              & for    & $j\neq v$    & $X_j(x)$                                       & for    & $j\neq v$   \\
                                & $X_v(z)$              & for    & $j=v$        & $Z_v(\sm x)$                                      & for    & $j=v$       \\ \hline
\multirow{2}{*}{$M_v(m^{\sm 1})$} & $Z_j(z)$              & for    & $j\neq v$    & $X_j(x)$                                       & for    & $j\neq v$   \\
                                & $Z_v(m z)$ & for    & $j=v$        & $X_v(m^{\sm 1} x)$                               & for    & $j=v$       \\ \hline
\multirow{2}{*}{$S_v(\lambda)$, $p$ odd} &
  \multirow{2}{*}{$Z_j(z)$} &
  \multicolumn{2}{l|}{\multirow{2}{*}{$\forall j$}} &
  $X_j(x)$ &
  for &
  $j\neq v$ \\
                                &                       & \multicolumn{2}{l|}{} & $\chi(2^{\sm 1}\lambda x^2) X_v(x)Z_v(\lambda x)$ & for    & $j=v$       \\ \hline
\multirow{2}{*}{$S_v$, $p$ even} &
  \multirow{2}{*}{$Z_j(z)$} &
  \multicolumn{2}{l|}{\multirow{2}{*}{$\forall j$}} &
  $X_j(x)$ &
  for &
  $j\neq v$ \\
                                &                       & \multicolumn{2}{l|}{} & $\chi_4(x^2) X_v(x)Z_v( x)$             & for    & $j=v$       \\ \hline
\multirow{2}{*}{$CX_{s\rightarrow t}$} &
  $Z_j(z)$ &
  for &
  $j\neq t$ &
  $X_j(x)$ &
  for &
  $j \neq s$ \\
                                & $Z_s(\sm z) Z_t(z)$      & for    & $j=t$        & $X_s(x)X_t(x)$                                 & for    & $j=s$       \\ \hline
\multirow{2}{*}{$Z_v(a)$} &
  \multirow{2}{*}{$Z_j(z)$} &
  \multicolumn{2}{l|}{\multirow{2}{*}{$\forall j$}} &
  $X_j(x)$ &
  for &
  $j\neq v$ \\
                                &                       & \multicolumn{2}{l|}{} & $\chi(ax)X_v(x)$                               & for    & $j=v$       \\ \hline
\multirow{2}{*}{$X_v(a)$} &
  $Z_j(z)$ &
  for &
  $j=v$ &
  \multirow{2}{*}{$X_j(x)$} &
  \multicolumn{2}{l}{\multirow{2}{*}{$\forall j$}} \\
                                & $\chi(\sm az)Z_v(z)$     & for    & $j\neq v$    &                                                & \multicolumn{2}{l}{} \\ \hline
\multirow{3}{*}{$CZ_{s\rightarrow t}$} &
  \multirow{3}{*}{$Z_j(z)$} &
  \multicolumn{2}{l|}{\multirow{3}{*}{$\forall j$}} &
  $X_j(x)$ &
  for &
  $j\neq s, t$ \\
                                &                       & \multicolumn{2}{l|}{} & $X_s(x)Z_t(x)$                                 & for    & $j=s$       \\
                                &                       & \multicolumn{2}{l|}{} & $X_t(x)Z_s(x)$                                 & for    & $j=t$       \\ \hline\hline
\end{tabular}
\caption{Commutation relations between the Weyl operators, $Z(z)$ and $X(x)$, and the Clifford generators (see Table~\ref{Table:1}).}
\label{Table:commutation_relations}
\end{table}

\subsection{\texorpdfstring{$Z$}{Z} measurement} \label{Sec.Proofs Measurements:Action Z}
For the following derivations we note that the computational basis is given as $\ket{b}=\ket{(1,0),b}$. First, we derive a useful expression for the projector $P(Z_v,b)$
\begin{equation}\label{equ:proj_z_equ}
\begin{aligned}
P(Z_v,b)&=\frac{1}{d}\sum_{y \in \Fd} \bar \chi(yb)Z_v(y)=\frac{1}{d}\sum_{y,k \in \Fd} \bar \chi(yb)\chi(yk) \ket{k}_v \bra{k} \\
&=\frac{1}{d}\sum_{k \in \Fd }\ket{k }_v\bra{k} \underbrace{\sum_{y \in \Fd} \bar \chi(y(b \sm k))}_{= d\delta_{b,k}} \\
&= \ket{b}_v \bra{b}.
\end{aligned}
\end{equation}
Equivalent to the definition of a graph state using stabilizers, Eq.~\eqref{Eq Graph}, a graph state $\ket{G}$ is given by $\ket{G}= \prod_{(i,j)\in E}CZ_{i\rightarrow j}(A_{ij}) \ket{+}^{\otimes V}$, where $\ket{+}:=\sum_{x \in \Fd} \ket{x}/\sqrt{d}$. By deferring the application of the controlled $Z$ gates for qudit $v$ yields
\begin{equation} \label{eq:graph_proc}
\ket{G}=\prod_{j \in V} CZ_{v \rightarrow j}(A_{vj}) \ket{+}_v \ket{G \setminus v},
\end{equation}
where we can write the $CZ$ term as
\begin{equation*}
\prod_{j \in V} CZ_{v \rightarrow j}(A_{vj})=\prod_{j \in V} \sum_{k \in \Fd} \ket{k}_v \bra{k} Z_{j}(k A_{vj})= \sum_{k \in \Fd} \ket{k}_v \bra{k} Z_{j}(k A_{v}).
\end{equation*}
Then, the projector expression Eq.~\eqref{equ:proj_z_equ} acting on $\ket{G}$ yields
\begin{align*}
P(Z_v,b) \ket{G} &= \ket{ b}_v \bra{b} \sum_{k \in \Fd} \ket{k}_v \bra{k} Z(kA_v) \ket{+}_v \ket{G \setminus v} \\
&= Z(bA_v) \ket{ b}_v \bra{b} \ket{+}_v \ket{G \setminus v} \\
&= \frac{1}{\sqrt d} \ket{ b}_v Z(b A_v) \ket{G \setminus v},
\end{align*}
which proves Eq.~\eqref{Eq.Zgraph}. The action of the projector on a state will, in general, produce a not normalized state, which can be accounted for via dividing by the square root of the probability of the measurement outcome occurring. Note that each outcome has the same probability $1/d$. This is implicitly done in the main text, where we present the actions of the projectors.

\subsection{\texorpdfstring{$Y$}{Y}-type measurement} \label{Sec.Proofs Measurements:Action Y}
For the following derivations we note that the $W(1,m)$ basis elements are $R(\sm m)\ket{b}=\ket{(1,m),b}$. We first show the local Clifford equivalence claimed in Eq.~\eqref{Eq.Yequ}, using Eq.~\eqref{Eq.local_compl} and Eq.~\eqref{Eq.projector},
\begin{align*}
L_v(\sm m) P(Z_v,b) L_v(m) &=\frac{1}{d} \sum_{y \in \Fd} L_v(\sm m) \bar \chi(yb) Z_v(y) L_v(m) \\
& \comrel \frac{1}{d} \sum_{y \in \Fd} \bar \chi(yb) \chi(\sm 2^{ \sm 1} m y^2)Z_v(y)X_v(my) \\
& =P(W_v(1,m),b).
\end{align*}
Now to show Eq.~\eqref{Eq.Ygraph}, we apply the projector onto a graph state and use Eq.~\eqref{Eq.Zgraph} and Eq.~\eqref{Eq.Yequ}
\begin{align*}
P(W_v(1,m),b) \ket{G} &=L_v(\sm m) P(Z_v,b) L_v(m) \ket{G} \\ 
& = L_v(\sm m) P(Z_v,b) \ket{\tau_v(m)(G)}\\
& = \frac{1}{\sqrt d} L_v(\sm m) \ket{b}_v Z(bmA_v)  \ket{\tau_v(m)(G) \setminus v}
\\
& = \frac{1}{\sqrt d} R_v(\sm m) \ket{b}_v S(m A_v) Z(bmA_v) \ket{\tau_v(m)(G) \setminus v},
\end{align*}
where one can identify $U_{(1,m),b}^{(v)}$ and $G'$.

\subsection{\texorpdfstring{$X$}{X} measurement}
For the following derivations we note that the $X(1)$ basis elements are $ S(\sm 1) R(\sm 1)\ket{b}=\ket{(0,1),b}$. The proof of the local Clifford equivalence of the $X$ projector and the $W(1,1)$ projector presented in Eq.~\eqref{Eq.Xequ}, is as follows [using Eq.~\eqref{Eq.local_compl} and Eq.~\eqref{Eq.projector}], where $w_0$ is a neighbor of $v$ and $r=\sm A_{w_0,v}^{\sm 2}$, 
\begin{align*}
L_{w_0}(\sm r)P(W_v(1,1),b)L_{w_0}(r) &= \frac{1}{d}\sum_{y \in \Fd} \bar \chi(by) \chi(\sm 2^{\sm 1}y^2) S_v(\sm 1) Z_v(y) X_v(y)S_v(1)\\
&\comrel \frac{1}{d}\sum_{y \in \Fd} \bar \chi(by) \chi(\sm 2^{\sm 1}(y^2 \sm y^2)) X_v(y) \\
&= P(X_v,b).
\end{align*}
To show Eq.~\eqref{Eq.Xgraph} we apply $P(X_v,b)$ on a graph state $\ket{G}$, using Eq.~\eqref{Eq.Ygraph} and Eq.~\eqref{Eq.Xequ}, yielding 
\begin{align*}
    P(X_v,b) \ket{G} &= L_{w_0}(\sm r) P(W_v(1,1),b) L_{w_0}(r) \ket{G} \\
    &= L_{w_0}(\sm r)  P(W_v(1,1),b) \ket{\tau_{w_0}(r)(G)}  \\
    &= \frac{1}{\sqrt d} L_{w_0}(\sm r) R_v(\sm 1)\ket{b}_v S(A'^2_v)Z(bA'_v)\ket{\tau_v(1)( \tau_{w_0}(r)(G) )\setminus v} \\
    &= \frac{1}{\sqrt{d}} S_v(r A_{w_0,v}^2) R_v(\sm 1) \ket{b}_v R_{w_0}(\sm r) S(r(A_{w_0}^2 \sm \bm e_v A^2_{w_0,v}))  S(A'^2_v)Z(bA'_v)\ket{\tau_v(1)( \tau_{w_0}(r)(G) )\setminus v}  \\
    &=\frac{1}{\sqrt{d}} S_v(\sm 1) R_v(\sm 1) \ket{b}_v R_{w_0}(\sm r)S(r(A_{w_0}^2 \sm \bm e_v A^2_{w_0,v}))  S(A'^2_v)Z(bA'_v)\ket{\tau_v(1)( \tau_{w_0}(r)(G) )\setminus v} \\
    &=\frac{1}{\sqrt{d}} S_v(\sm 1) R_v(\sm 1) \ket{b}_v R_{w_0}(\sm r) S(r(\tilde{A}_{w_0}^2))  S(A'^2_v)Z(bA'_v)\ket{\tau_v(1)( \tau_{w_0}(r)(G) )\setminus v},
\end{align*}
where $A'$ is the adjacency matrix of $\tau_{w_0}(r)(G)$, $\tilde{A}_{w_0}^2=A_{w_0}^2 \sm \bm e_v A^2_{w_0,v}$, and one can identify $U^{(v)}_{(0,1),b}$ and $G'$. 

\subsubsection{Choice of neighbor \texorpdfstring{$w_0$}{w0}}
The choice of the neighbor $w_0$ for the $X$ measurement is arbitrary. Here, we show that the resulting graph state obtained from a different neighbor $w_1$ is local Clifford equivalent to the one obtained from $w_0$. This can be readily obtained after using the Clifford equivalence from Eq.~\eqref{Eq.Xequ} for neighbor $w_1$, the Clifford equivalence from Eq.~\eqref{Eq.Yequ}, and the action of the $X$ measurement on a graph state Eq.~\eqref{Eq.Xgraph}.
\begin{align*}
\ket{\tau_v(1)(\tau_{w_0}(r)(G)) \setminus v}&= \bra{(0,1),b}_v U_{(0,1),b}^{(v) \dag} P(X_v,b) \ket{G} \\
&= \bra{ b}_v Z(\sm bA'_v) \tilde{L}_v(1) L_{w_0}(r) P(X_v,b)\ket{G} \\
&= \bra{ b}_v Z(\sm bA'_v) \tilde{L}_v(1) L_{w_0}(r)  L_{w_1}(\sm \tilde{r}) \tilde{\tilde{L}}_v(\sm 1)P(Z_v,b)  L_{w_1}( \tilde{r}) \tilde{\tilde{L}}_v(1) \ket{G}  \\
&= \bra{ b}_v Z( \sm bA'_v) \tilde{L}_v(1) L_{w_0}(r)  L_{w_1}(\sm \tilde{r}) \tilde{\tilde{L}}_v(\sm 1)P(Z_v,b)  \ket{\tau_v(1)(\tau_{w_1}(\tilde{r})(G))} \\
&= \bra{ b}_v Z(\sm bA'_v) \tilde{L}_v(1) L_{w_0}(r)  L_{w_1}(\sm \tilde{r}) \tilde{\tilde{L}}_v(\sm 1)Z( b\tilde{A'}_v) \ket{ b}_v \ket{\tau_v(1)(\tau_{w_1}(\tilde{r})(G)) \setminus v} \\
&=Z(\sm bA'_v) S(\sm A'_v) R_{w_0}(r)S(\sm A_{w_0}r\sm \bm e_v)R_{w_1}(\sm \tilde{r}) S(\tilde{r}A_{w_1}+ \bm e_v)S(\tilde{A'}_v) Z( b\tilde{A'}_v)\ket{\tau_v(1)(\tau_{w_1}(\tilde{r})(G)) \setminus v},
\end{align*}
where $\tilde{L}$ is the local complementation defined with adjacency matrix $A'$ of $\tau_{w_0}(r)(G)$, $\tilde{r}=\sm A^{\sm 2}_{w_1,v}$, $\tilde{A'}$ is the adjacency matrix of $\tau_{w_1}(\tilde{r})(G)$, and $\tilde{\tilde{L}}$ is the local complementation defined with adjacency matrix $\tilde{A'}$.

\subsection{\texorpdfstring{$X(m)$}{X(m)} measurement}
For the following derivations we note that the $X(m)$ basis elements are $M(m) S(\sm 1) R(\sm 1)\ket{b}=\ket{(0,m),b}$. We begin with showing the Clifford equivalence given in Eq.~\eqref{Eq.Xmequ}, using Eq.~\eqref{Eq.projector}, 
\begin{align*} 
M_v(m)P(X_v,b) M_v(m^{\sm 1})&=\frac{1}{d}\sum_{y \in \Fd} \bar \chi(by) M_v(m) X_v(y)M_v(m^{\sm 1})\\ 
&\comrel \frac{1}{d} \sum_{y \in \Fd} \bar \chi(by) X_v(my) \\
& = P(X_v(m),b). 
\end{align*}
In order to show Eq.~\eqref{Eq.Xmgraph} we apply the $P(X_v(m),b)$ on a graph state $\ket{G}$, using Eq.~\eqref{Eq.Xgraph} and Eq.~\eqref{Eq.Xmequ},
\begin{align*}
    P(X_v(m),b) \ket{G} &= M_v(m) P(X_v,b) M_v(m^{\sm 1}) \ket{G} \\
    &=M_v(m) P(X_v,b) \ket{G \circ_m v } \\
    &=\frac{1}{\sqrt{d}} M_v(m) S_v(\sm 1) R_v(\sm 1) \ket{b}_v R_{w_0}(\sm r) S(r(\tilde{A}_{w_0}^2))  (S(mA'^2_v)Z(bmA'_v))\ket{\tau_v(1)( \tau_{w_0}(r)(G \circ_m v) )\setminus v},
\end{align*}
where one can identify $U^{(v)}_{(0,m),b}$ and $G'$.

\subsection{\texorpdfstring{$Z(n)$}{Z(n)} measurement}
For the following derivations we note that the $Z(n)$ basis elements are $ M(n^{\sm 1}) \ket{b}=\ket{(n,0),b}$. First, we show the Clifford equivalence from Eq.~\eqref{Eq.Znequ}, using Eq.~\eqref{Eq.projector},
\begin{align*} 
M_v(n^{\sm 1})P(Z_v,b) M_v(n)&=\frac{1}{d} \sum_{y \in \Fd} \bar \chi(by) M_v(n^{\sm 1}) Z_v(y)M_v(n)\\ 
& \comrel \frac{1}{d} \sum_{y \in \Fd} \bar \chi(by) Z_v(ny) \\
& = P(Z_v(n),b). 
\end{align*}
For completeness we show the action of $P(Z_v(n),b)$ on a graph state $\ket{G}$, although this is contained in $P(W(n,m), b)$,
\begin{align*}
P(Z_v(n),b) \ket{G} &= M_v(n^{\sm 1}) P(Z_v,b) M_v(n) \ket{G} \\
&= M_v(n^{\sm 1}) P(Z_v,b) \ket{G \circ_{n^{ \sm 1}} v} \\
&= \frac{1}{\sqrt{d}} M_v(n^{\sm 1}) \ket{b}_v Z(bn^{ \sm 1} A_v ) \ket{(G \circ_{n^{ \sm 1}} v) \setminus v }.
\end{align*}
\subsection{\texorpdfstring{$W(n,m)$}{W(n,m)} measurement}
For the following derivations we note that the $W(n,m)$ basis elements are $R(\sm m/n) M(n^{\sm 1})\ket{b}=\ket{(n,m),b}$. We first show the Clifford equivalence claimed in Eq.~\eqref{Eq.Wnmequ}, using Eq.\eqref{Eq.projector},
\begin{align*}
L_v(\sm m/n)P(Z_v(n),b) L_v(m/n)&=\frac{1}{d}\sum_{y \in \Fd} \bar \chi(bw) L_v(\sm m/n)Z_v(ny) L_v(m/n) \\
& \comrel \frac{1}{d}\sum_{y \in \Fd} \bar \chi(by)\chi(\sm 2^{\sm 1}mny^2) Z_v(ny)X_v(my) \\
& = P(W_v(n,m),b).
\end{align*}
To show Eq.~\eqref{Eq:xmznequ}, we apply $P(W_v(n,m),b)$ onto a graph state $\ket{G}$, using Eq.\eqref{Eq.Zgraph}, Eq.~\eqref{Eq.Znequ} and Eq.~\eqref{Eq.Wnmequ}, 
\begin{align*}
    P(W_v(n,m),b) \ket{G}&= L_v(\sm m/n) P(Z(n),b) L_v(m/n) \ket{G} \\
   &=\frac{1}{\sqrt{d}} L_v(\sm m/n) P(Z(n),b) \ket{\tau_v(m/n)(G)} \\
    &=\frac{1}{\sqrt{d}} L_v(\sm m/n) M_v(n^{\sm 1}) \ket{ b}_v Z(b n^{\sm 1} A') \ket{\tau_v(m/n)(G)\circ_{n^{\sm 1}} v \setminus v} \\
    &=\frac{1}{\sqrt{d}} R_v(\sm m/n) M_v(n^{\sm 1}) \ket{b}_v S(m/n A^2_v)Z(b n^{\sm 1} A') \ket{\tau_v(m/n)(G)\circ_{n^{\sm 1}} v \setminus v}, \\
\end{align*}
where $A'$ is the adjacency matrix of the graph $\tau_v(m/n)( G )$, and one can identify $U^{(v)}_{(n,m),b}$ and $G'$.

\section{Proofs for the update rules} \label{Sec.Proofs:Update rules}
Here we first derive the building-block update rules presented in Sec.~\ref{Sec.Update rules}, where we omit phases as they do not contribute in Pauli-diagonal channels. We also give the update rules for all the Weyl measurement operations.

\subsection{Building blocks}
\subsubsection{Local multiplication}
The update rule for $M(m^{\sm 1})$, given in Eq.~\eqref{Eq.ur_local_mult}, directly follows from the commutation relation presented in Table~\ref{Table:commutation_relations}. 

\subsubsection{Local complementation}
The update rule for $L(m)$, given in Eq.~\eqref{EqLCupdate}, can be readily obtained via recalling the definition of the local complementation given in Eq.~\eqref{Eq.local_compl} and the commutation relations in  Table~\ref{Table:commutation_relations}. If the operators act on different vertices, then one notices that they commute, since $[S(\lambda),Z(z)]=0$. Now consider that the local complementation and a noise operator $Z(z)$ act on the same vertex,
\begin{align*}
    L_v(m)Z_v(z)&= H_v S_v(m) H_v^\dag \bigotimes_{j } S_j(\sm A_{v,j}^2m) Z_v(z) \\
    &= H_v S_v(m) H_v^\dag Z_v(z) \bigotimes_{j } S_j(\sm A_{v,j}^2m) \\
    &\comrel H_v S_v(m) X_v(z) H_v^\dag \bigotimes_{j } S_j(\sm A_{v,j}^2m) \\
    &\comrel H_v X_v(z)Z_v(mz) S_v(m) H_v^\dag \bigotimes_{j } S_j(\sm A_{v,j}^2m) \\
    &\comrel Z_v(z) X_v(\sm z m) H_v S_v(a) H_v^\dag \bigotimes_{j } S_j(\sm A_{v,j}^2m) \\
    &=Z_v(z) X_v(\sm z m) L_v(m),
\end{align*}
and if they act on a graph state $\ket{G}$, then we obtain 
\begin{align*}
     L_v(m)Z_v(z) \ket{G}&= Z_v(z) X_v(\sm z m) L_v(m) \ket{G} \\
     &=Z_v(z) X_v(\sm z m) \ket{\tau_v(m)(G)} \\
     &=Z_v(z) Z(m z A_v) \ket{\tau_v(m)(G)}.
\end{align*}

\subsubsection{\texorpdfstring{$Z$}{Z} measurement}
The update rule for the $Z$ measurement, given in Eq.~\eqref{Eq.ur_z_meas}, can be obtained via the definition of the $Z$ measurement operation, Eq.~\eqref{Eq.Zgraph}. If they act on different vertices, then they commute since $Z$ operators commute with the $Z$ projector and other $Z$ operators. However, the case where they act on the same vertex is as follows: 
\begin{align*}
    \mathcal{L} (Z_v,b) Z_v(z) &= \bra{(1,0), b}_v Z(b A_v) P(W_v(1,0),b)  Z_v(z) \\
    &= \bra{(1,0), b}_v Z(b A_v) Z_v(z) P(W_v(1,0),b)  \\
    &= \bra{(1,0), b}_v Z_v(z) Z(b A_v)  P(W_v(1,0),b)  \\
    &= \bra{(1,0), b}_v Z(b A_v)  P(W_v(1,0),b)  \\
    &=  \mathcal{L} (Z_v,b).
\end{align*}
 
\subsection{Weyl measurement operators}
In the following, we make use of the fact that each Weyl projector is Clifford equivalent to the $Z$ projector. So we can write the projection of an Weyl operator $O$ on a graph state $\ket{G}$ always as $P(O_v,b)\ket{G} = U^\dag P(Z_v,b) U \ket{G}$, where $U$ is said Clifford. Applying this projector onto a graph state, we obtain 
\begin{equation*}
    P(O_v,b)\ket{G} = U^\dag P(Z_v,b) U \ket{G}=U^\dag P(Z_v,b)  \ket{G'}= U^\dag Z(bA'_v) \ket{(1,0), b}_v \ket{G' \setminus v},
\end{equation*}
So we identify the measurement operator as $\mathcal{L}(O_v,b)=\bra{(1,0), b}_v Z(\sm bA_v) U P(O_v,b)$. Now we consider the general scenario where a noise operator is acting on qudit $u$ of the graph state
\begin{equation*}
    \mathcal{L} (O_v,b) Z_u(z) \ket{G}= \bra{(1,0), b}_v Z(\sm b A'_v) P(Z_v,b) U Z_u(z) \ket{G} = \mathcal{L} (Z_v, b) U Z_u(z)\ket{G}.
\end{equation*}
Using the update rules yields 
\begin{align*}
    \mathcal{L} (O_v,b) Z_u(z) \ket{G}& =\mathcal{L} (Z_v, b) U Z_u(z) \ket{G} \\
    &=\mathcal{L} (Z_v, b)  \tilde{Z}_u(z) U \ket{G} \\
    &= \tilde{\tilde{Z}}_u(z) \ket{G' \setminus v},
\end{align*}
where $\tilde{\tilde{Z}}_u(z)$ is the updated noise operator, after applying the update rules corresponding to the Clifford $U$ and the $Z$ measurement operator. Therefore, the update rule for a general local Weyl measurement is the concatenation of the update rules of the Clifford transformation $U$ with the $Z$ measurement. 

\subsubsection{\texorpdfstring{$Y$}{Y}-type measurement}
From Eq.~\eqref{Eq.Yequ} we find that for $\mathcal{L} (W_v(1,m),b)$ the corresponding Clifford is $U=L_v(m)$. Using the update rules of the local complementation, Eq.~\eqref{EqLCupdate}, and the $Z$ measurement, Eq.~\eqref{Eq.ur_z_meas}, we find that if $\mathcal{L} (W_v(1,m),b)$ and $Z_u(z)$ act on different vertices, then they commute, but if they act on the same vertex, then the overall update rule yields
\begin{equation}
    \mathcal{L} (W_v(1,m),b) Z_v(z) \ket{G}= Z(zmA_v) \mathcal{L}(W_v(1,m),b) \ket{G}.
\end{equation}

\subsubsection{\texorpdfstring{$X$}{X} measurement}
From Eq.~\eqref{Eq.Yequ} and Eq.~\eqref{Eq.Xequ} we gather the corresponding Clifford as $U=L_v'(1) L_{w_0}(r)$, where $L_v'(1)$ is the local complementation of $G'=\tau_{w_0}(r)(G)$. Using the update rules for the local complementation, Eq.~\eqref{EqLCupdate}, and for the $Z$ measurement, one observes that $\mathcal{L}(X_v,b)$ and $Z_u(z)$ commutes if $u\neq v$ and $u\neq w_0$; however, if the noise operator acts on the measurement vertex $v$, then the total update rule yields
\begin{equation}
    \mathcal{L}(X_v,b) Z_v(z)\ket{G}= Z_v(zA'_v)\mathcal{L}(X_v,b)\ket{G},
\end{equation}
where $A'$ is adjacency matrix of $G'$. If the noise operator acts on the neighbor vertex $w_0$, then the total update rule yields
\begin{equation}
    \mathcal{L}(X_v,b) Z_{w_0}(z) \ket{G} = Z_{w_0}(z)Z(zr(A_{w_0} \sm \bm{e}_v A_{w_0,v})) Z(zr A_{w_0,v} A'_{v}) \mathcal{L}(X_v,b)  \ket{G}.
\end{equation}

\subsubsection{\texorpdfstring{$X(m)$}{X(m)} measurement}
From Eq.~\eqref{Eq.Yequ}, Eq.~\eqref{Eq.Xequ}, and Eq.~\eqref{Eq.Xmequ} we gather the corresponding Clifford as $U=L_v''(1) L'_{w_0}(r) M_v(m^{\sm 1})$, where $L'_{w_0}(r)$ is the local complementation of $G'=G \circ_m v$ and $L_v''(1)$ is the local complementation of $G''=\tau_{w_0}(r)(G \circ_m v)$.
Using the update rules for the local complementation, Eq.~\eqref{EqLCupdate}, for the local multiplication, Eq.~\eqref{Eq.ur_local_mult}, and for the $Z$ measurement, Eq.~\eqref{Eq.ur_z_meas}, we find that if $\mathcal{L}(X_v(m),b)$ and $Z_u(z)$ do not act on the same vertex nor the noise operator acts on the neighbor $w_0$, then they commute; however, if they act on the same vertex, then the overall update rule yields
\begin{equation}
    \mathcal{L}(X_v(m),b) Z_v(z)\ket{G}= Z_v(zmA''_v)\mathcal{L}(X_v(m),b)\ket{G},
\end{equation}
where $A''$ is adjacency matrix of $G''$. If the noise operator acts on the neighbor $w_0$, then the total update rule yields
\begin{equation}
    \mathcal{L}(X_v(m),b) Z_{w_0}(z) \ket{G} = Z_{w_0}(z)Z(zr(A'_{w_0} \sm \bm{e}_v A'_{w_0,v})) Z(zr A'_{w_0,v} A''_{v}) \mathcal{L}(X_v(m),b)  \ket{G},
\end{equation}
where $A'$ is the adjacency matrix of $G'$.

\subsubsection{\texorpdfstring{$Z(n)$}{Z(n)} measurement}
From Eq.~\eqref{Eq.Znequ} we find that for $\mathcal{L} (Z_v(n),b)$ the corresponding Clifford is $U=M_v(n)$. Using the update rules of the local multiplication, Eq.~\eqref{Eq.ur_local_mult}, and the $Z$ measurement, Eq.~\eqref{Eq.ur_z_meas}, we find that if $\mathcal{L}(W_v(n,0),b)$ and $Z_u(z)$ act on different vertices, then they commute; however, if they act on the same vertex, then the total update rule yields
\begin{equation}
    \mathcal{L} (Z_v(n),b)Z_v(z) \ket{G}= \mathcal{L} (Z_v(n),b) \ket{G}.
\end{equation}

\subsubsection{\texorpdfstring{$W(n,m)$}{W(n,m)} measurement}
From Eq.~\eqref{Eq.Znequ} and Eq.~\eqref{Eq.Wnmequ} we find the corresponding Clifford is $U=M_v(n) L_v(m/n)$. Using the update rules for the local complementation, Eq.~\eqref{EqLCupdate}, the $Z$ measurement, Eq.~\eqref{Eq.ur_z_meas}, and the local multiplication, Eq.~\eqref{Eq.ur_local_mult}, we observe that $\mathcal{L}(W_v(n,m),b)$ commutes with $Z_u(z)$ if they act on different vertices, but if they act on the same vertex, then the overall update rule yields
\begin{equation}
    \mathcal{L} (W_v(n,m),b) Z_v(z)\ket{G}= Z\left(z\frac{m}{n} A_v \right) \mathcal{L} (W_v(n,m),b) \ket{G}.
\end{equation}

\section{Qudit Noisy Graph State Formalism for arbitrary finite dimensions} \label{Sec.Qudit graph state formalism for arbitrary finite dimensions}
Here, we show how one can obtain a noisy stabilizer formalism for qudit graph states of arbitrary finite dimension. To this end, we use the linearized stabilizer formalism introduced in \cite{2-2013}. In contrast to before, we work no longer over finite fields of size $d=p^m$ but over finite rings $\mathbbm{Z}_d$ of size $d$. The computational basis states $\ket{i}$ are labeled with integers out of $\mathbbm{Z}_d$. We define the generalized Pauli $X$ operator and the generalized Pauli $Z$ operator via the action on a basis state $\ket{x}$ 
\begin{equation}
    \begin{aligned}
         X \ket{x} &= \ket{x+1 }, \\
         Z \ket{x} &= \omega^{x} \ket{x},
        \end{aligned}
\end{equation}
where $\omega=\exp(2\pi i /d)$ and the addition in the $X$ gate definition is the addition over $\mathbbm{Z}_d$. We define $\tau=( \sm 1)^d \exp( \pi i /d)$ such that $\tau^2=\omega$. With this the $d$-dimensional Pauli group $\mathcal{P}_d$ is defined via $\mathcal{P}_d=\langle \tau \mathbbm{1}, X, Z \rangle$. The Weyl operators $W(z,x)$ are defined via 
\begin{equation}
    W(z,x)=\tau^{\sm zx} Z^zX^x.
\end{equation}
The $n$-qudit Weyl operator is given by 
\begin{equation}
    W(\bm z, \bm x)=\tau^{\sm \bm z \bm x} Z^{\bm z}X^{\bm x}=\tau^{\sm  \bm {z  x}}\bigotimes_v Z_v^{ z_v} \bigotimes_v X_v^{ x_v},
\end{equation}
where $\bm z, \bm x \in \mathbbm{Z}_d^n$,  $z_v \in \bm z$, and $x_v \in \bm x$. One defines the single-qudit Clifford operators
\begin{equation}
    S=\sum_{x \in \mathbbm{Z}_d} \tau^{x^2} \ketbra{x}, \qquad 
    H=\sum_{x,y \in \mathbbm{Z}_d} \tau^{2xy} \ket{x}\bra{y}, \qquad
    M_a= \sum_{x \in \mathbbm{Z}_d}  \ket{ax}\bra{x}, \ a \in \mathbbm{Z}_d^*, 
\end{equation}
where $ \mathbbm{Z}_d^*$ is the group of the multiplicative units in $\mathbbm{Z}_d$, i.e., those elements with multiplicative inverses. The controlled $X$ and the controlled $Z$ operators are defined as follows
\begin{equation}
        CX= \sum_{x,y \in \mathbbm{Z}_d} \ketbra{x} \ket{x+y}\bra{y}, \qquad
        CZ=\sum_{x,y \in \mathbbm{Z}_d} \tau^{2xy}\ketbra{x,y}.
\end{equation}
The Clifford operators $S$, $H$, $CZ$, and $M_a$ (where $a$ is ranging over the units of $\mathbbm{Z}_d$) generate the symplectic Clifford group \cite{2-2013}. Using the here introduced definition of the Weyl operators, one observes the same commutation relation between the Clifford generators and the Pauli $X$ and $Z$, as the ones in the prime-power case, shown in Table~\ref{Table:commutation_relations}. 

We define a qudit graph state associated to a weighted graph $G=(V,E,A)$ (where the edge weights are now integers out of $\mathbbm{Z}_d$) as the $+1$ eigenstate of the eigenvalue equation
\begin{equation}
    X_v Z^{A_v} \ket{G}=\ket{G},\quad \forall v \in V.
\end{equation}
From this, one observes that one can translate the action of an $X_v$ operator on a graph state $\ket{G}$ to powers of $Z$ on the neighborhood of $v$
\begin{equation}
    X_v \ket{G}= X_v (X_v Z^{A_v})^{d \sm 1}\ket{G}=Z^{(d \sm 1)A_v} \ket{G}= Z^{\sm A_v} \ket{G}.
\end{equation}
The update rules for the (symplectic) Clifford operators are also equivalent to the ones in the prime-power case, since the commutation relations for (symplectic) Clifford and Weyl operators are equivalent to the prime-power case (they map equivalent Weyl operators to equivalent Weyl operators under commutation) and the translation of general Weyl operator on graph states is equivalent to the one in the prime-power case. 

The projector of a Pauli operator $O \in \mathcal{P}_d$ associated to measurement outcome $h$ is defined as $P(O,h)=\frac{1}{d}\sum_{j \in \mathbbm{Z}_d} \tau^{\sm 2hj} O^j$. This definition is formally equivalent to the one presented in the prime-power case after the identification of $\chi(x) \sim \omega^x$ and $  W(t(a,b)) \sim W(a,b)^t$. Observing this formal equivalence, one can derive equivalent expressions for the projectors and their action on the graph states, as presented in Appendix~\ref{Sec.Proofs Measurements}, provided the Clifford equivalences for the projectors do not require multiplicative inverses of zero-divisors. Then, the update rules for the measurements are also equivalent to the prime-power case, provided, once again, the Clifford equivalences for the projectors do not require multiplicative inverses of zero-divisors. 

\section{Proofs of Sec.~\ref{Sec.Application of method}} \label{Sec.Proofs: Linear chain}
Here we prove the results presented in Sec.~\ref{Sec.Application of method} using the update rules (u.r.) presented in Appendix~\ref{Sec.Proofs:Update rules}.

\subsection{Updated final noise maps} \label{SubSec.Form of final noise maps}
Here we show the form of the updated noise channels acting on the generalized Bell pair. The procedure to obtain the generalized Bell pair from the linear cluster state is to measure qudits 2 to $N \sm 1$ in the $W(1,1)$ basis. We encode the order of these measurements with a vector of length $n:= N \sm 2$, $\bm \sigma= (\sigma_1, \sigma_2, \dots, \sigma_n)$, where $\sigma_i$ takes the value of the label of a qudit from 2 to $N \sm 1$. Note that for the vector to match the way quantum operations are applied, the first measurement is performed on qudit $\sigma_n$, and the last on $\sigma_1$. Thus, the overall manipulation operator for a given sequence of measurements is 
\begin{equation*}
\mathcal{L}_{\bm \sigma}:=\mathcal{L}(W_{\sigma_1}(1,1),b)\cdots \mathcal{L}(W_{\sigma_n}(1,1),b).
\end{equation*}

\subsubsection{Updated noise maps of edge qudits}
We begin by showing the form for the noise map on qudits 1 and $N$. First, we define $I=(1,\dots, N)$ as the initial ordered vertex label set, and $I^{\sigma_i}=I \setminus \{\sigma_i, \dots, \sigma_n \}$ as the ordered vertex label set after partial measurement sequence $\mathcal{L}(W_{\sigma_i}(1,1),b)\cdots \mathcal{L}(W_{\sigma_n}(1,1),b)$, where $I^{\sigma_i}(j)$ denotes the $j$th element of the ordered set. Moreover, the graph state resulting from the partial measurement sequence until label $\sigma_i$ is denoted as
\begin{equation*}
    \ket{G^{\sigma_i}}:= \mathcal{L}(W_{\sigma_i}(1,1),b)\cdots \mathcal{L}(W_{\sigma_n}(1,1),b) \ket{G},
\end{equation*}
with $A^{\sigma_i}$ being the corresponding adjacency matrix. Due to symmetry, it is sufficient to analyze the noise map acting on qudit 1, as the final updated noise map obtained from the noise map acting on qudit $N$ is the same. Taking a general noise term from the depolarizing channel of qudit 1, Eq.~\eqref{Eq.Depol}, $N_1=Z_1(z)Z(\sm A_1x)=Z_1(z) Z_2(\sm x)$ we have
\begin{align*}
\mathcal{L}_{\bm \sigma}N_1 \ket{G}&=\mathcal{L}(W_{\sigma_1}(1,1),b) \cdots \mathcal{L}(W_{\sigma_n}(1,1),b) Z_1(z) Z_2(\sm x) \ket{G} \\
&= \mathcal{L}(W_{\sigma_1}(1,1),b) \cdots \mathcal{L}(W_{\sigma_j=2}(1,1),b) Z_1(z) Z_2(\sm x) \mathcal{L}(W_{\sigma_{j+1}}(1,1),b) \dots \mathcal{L}(W_{\sigma_n}(1,1),b) \ket{G} \\
&= \mathcal{L}(W_{\sigma_1}(1,1),b) \cdots \mathcal{L}(W_{\sigma_j=2}(1,1),b)Z_1(z) Z_2(\sm x)\ket{G^{\sigma_{j+1}}}  ,
\end{align*}
where we used that $N_1$ and $\mathcal{L}(W_{v}(1,1),b)$ commutes unless $v=2$. Using the update rule for the $\mathcal{L}(W(1,1), b)$ we find
\begin{align*}
\mathcal{L}_{\bm \sigma}N_1 \ket{G}&\uprule  \mathcal{L}(W_{\sigma_1}(1,1),b) \cdots \mathcal{L}(W_{\sigma_{j \sm 1}}(1,1),b) Z_1(z) Z(\sm x A^{\sigma_{j+1}}_{2}) \mathcal{L}(W_{2}(1,1),b) \ket{G^{\sigma_{j+1}}} \\
&=\mathcal{L}(W_{\sigma_1}(1,1),b) \cdots \mathcal{L}(W_{\sigma_{j \sm 1}}(1,1),b) Z_1(z \sm x) Z_{I^{\sigma_j=2}(2)}(\sm x) \mathcal{L}(W_{2}(1,1),b) \ket{G^{\sigma_j=2}},
\end{align*}
where we have used that the neighborhood of qudit 2 is formed by qudit 1 and the second element of $I^{\sigma_j=2}$. Using the fact that the measurement $\mathcal{L}(W_v(1,1),b)$ and the noise operator $Z_1(z\sm x) Z_{I^{\sigma_j=2}(2)}(\sm x)$ commute unless $v=I^{\sigma_j=2}(2)$ and the update rules, we find
\begin{align*}
\mathcal{L}_{\bm \sigma}N_1 \ket{G}&=\mathcal{L}(W_{\sigma_1}(1,1),b) \cdots \mathcal{L}(W_{\sigma_l=I^{\sigma_j=2}(2)}(1,1),b) Z_1(z \sm x) Z_{I^{\sigma_j=2}(2)}(\sm x) \mathcal{L}(W_{\sigma_{l+1}}(1,1),b) \cdots \mathcal{L}(W_{\sigma_{j \sm 1}}(1,1),b) \ket{G^{\sigma_j=2}} \\
&= \mathcal{L}(W_{\sigma_1}(1,1),b) \cdots \mathcal{L}(W_{\sigma_l=I^{\sigma_j=2}(2)}(1,1),b) Z_1(z \sm x) Z_{I^{\sigma_j=2}(2)}(\sm x)  \ket{G^{\sigma_{l+1}}} \\
&\uprule \mathcal{L}(W_{\sigma_1}(1,1),b) \cdots \mathcal{L}(W_{\sigma_{l\sm 1}}(1,1),b) Z_1(z \sm x) Z(\sm x A_{I^{\sigma_j=2}(2)})  \ket{G^{\sigma_{l}}} \\
&= \mathcal{L}(W_{\sigma_1}(1,1),b) \cdots \mathcal{L}(W_{\sigma_{l\sm 1}}(1,1),b) Z_1(z \sm 2 x) Z_{I^{\sigma_l}(2)}(\sm x)\ket{G^{\sigma_{l}}}. 
\end{align*}
We perform this procedure inductively and obtain
\begin{equation*}
\mathcal{L}_{\bm \sigma}N_1 \ket{G} = Z_1(z \sm \alpha x)Z_N(\sm x) \ket{G^{\sigma_1}},
\end{equation*}
with $\alpha \in \Fp$, where note that $\ket{G^{\sigma_1}}=\ket{G'}$ corresponds to the final graph state. Therefore, the updated noise map is 
\begin{equation*}
\widetilde{\mathcal{E}_1}=\lambda \rho' +\frac{1\sm \lambda}{d^2} \sum_{z,x \in \Fd} Z_1(z \sm \alpha x)Z_N(\sm x) \rho' (Z_1(z \sm \alpha x) Z_N(\sm x))^\dag,
\end{equation*}
which corresponds to Eq.~\eqref{eq.mtarget} after the redefinition $z=z\sm\alpha x$ and $x=\sm x$. 
\subsubsection{Updated noise maps of middle qudits}
Now we want to show that the updated noise maps of the middle vertices have the claimed form given in Eq.~\eqref{eq.minner}. We show this via induction over the number of middle qudits $n=N\sm 2$. The induction hypothesis (IH) is given a general noise term from the depolarizing channel, Eq.~\eqref{Eq.Depol}, $N_j=Z_j(z)Z(\sm A_j x)$ and a measurement strategy $\mathcal{L}_{\bm \sigma}$ the updated noise operator is 
\begin{equation*}
\text{(IH)}: \mathcal{L}_{\bm \sigma}N_j\ket{G} = Z_1(\alpha(n)(a(n)z \sm x))Z_N(\beta(n)(a(n)z \sm x)) \mathcal{L}_{\bm \sigma}\ket{G},
\end{equation*}   
where $\alpha(n)$, $\beta(n)$, $a(n)$, and $b(n)$ are integers from $\Fp$ which depend on the measurement order, the noise map of origin, and $n$. First, we prove the induction beginning (IB), where we consider a three-qudit linear cluster such that the number of middle qudits is $n=1$ and the measurement strategy is simply $\mathcal{L}_{\bm \sigma}=\mathcal{L}(W_{2}(1,1),b)$ and the noise operator is $N_2=Z_2(z) Z_1(\sm x) Z_N(\sm x)$
\begin{align*}
\text{(IB): } \mathcal{L}_{\bm \sigma} N_2\ket{G} \uprule Z_1(z \sm x) Z_N(z \sm x) \mathcal{L}_{\bm \sigma}\ket{G}.
\end{align*}
Then we can proceed with the induction step (IS), $n \rightarrow n+1$, such that the linear cluster now has $N+1$ qudits, with the edge qudits being 1 and $N+1$ and the middle qudits going from 2 to $N$, and the measurement vector $\bm\sigma$ is now of length $n+1$. Thus, we aim to prove
\begin{align*}
\text{(IS): }\mathcal{L}_{\bm \sigma}N_j \ket{G}&=  Z_1(\alpha(n+1)(a(n+1)z \sm x))Z_{N+1}(\beta(n+1)(a(n+1)z \sm x)) \mathcal{L}_{\bm \sigma}\ket{G}.
\end{align*}
Depending on where we act with the new measurement operator $\mathcal{L}(W_{N}(1,1),b)$ in our measurement order, three cases arise. Notice that $\mathcal{L}(W_{N}(1,1),b) $ commutes with $\mathcal{L}(W_j(1,1),b)$, for $j<N \sm 1$. 

\begin{itemize}
    \item\textbf{Case 1. }The measurement added in the induction step, $\mathcal{L}(W_{N}(1,1),b)$, is the last measurement being performed. The other measurements act before following the same order as in the induction hypothesis (IH). Thus, we can write the measurement process as $\bm\sigma = (N, \sigma_1, \dots, \sigma_n)$, where the last elements correspond to the $\bm\sigma$ of the (IH) scenario.
    
    Here we use the graph state definition introduced in Eq.~\eqref{eq:graph_proc} such that $\ket{G}_{V(N+1)} = CZ_{N \rightarrow N+1}  \ket{G}_{V(N)} \ket{+}_{N+1}$, where $V(m)$ denotes the vertices of a linear chain of length $m$. Also using this notation we define $\ket{G'}_{V(N)} = \mathcal{L}(W_{\sigma_1}(1,1),b) \cdots \mathcal{L}(W_{\sigma_n}(1,1),b)\ket{G}_{V(N)}$. Depending on which qudit $j\in\{2, \dots, N\}$ the noise operator $N_j$ is acting on, we get the following two cases:    \begin{itemize}
        \item[$\blacksquare$] For $j<N$:
    \begin{align*}
    \mathcal{L}_{\bm \sigma}N_j \ket{G}_{V(N+1)}&=\mathcal{L}(W_{N}(1,1),b)\mathcal{L}(W_{\sigma_1}(1,1),b) \cdots \mathcal{L}(W_{\sigma_n}(1,1),b) N_j CZ_{N \rightarrow N+1} \ket{G}_{V(N)}\ket{+}_{N+1} \\
    &\comrel \mathcal{L}(W_{N}(1,1),b) CZ_{N \rightarrow N+1} \mathcal{L}(W_{\sigma_1}(1,1),b) \dots \mathcal{L}(W_{\sigma_n}(1,1),b) N_j \ket{G}_{V(N)} \ket{+}_{N+1} \\
    &\overset{\text{(IH)}}{=} \mathcal{L}(W_{N}(1,1),b) CZ_{N \rightarrow N+1} Z_1(\alpha(n)(a(n)z \sm x))Z_{N}(\beta(n)(a(n)z \sm x)) \ket{G'}_{V(N)} \ket{+}_{N+1} \\
    & \comrel \mathcal{L}(W_{N}(1,1),b)  Z_1(\alpha(n)(a(n)z \sm x))Z_{N}(\beta(n)(a(n)z \sm x)) \ket{G}_{V(N+1)} \\
    &\uprule  Z_1((\alpha(n)+\beta(n))(a(n)z \sm x))Z_{N+1}(\beta(n)(a(n)z \sm x)) \mathcal{L}_{\bm\sigma} \ket{G}_{V(N+1)},
    \end{align*}
    where to prove (IS) we define $\alpha(n+1)=\alpha(n)+\beta(n)$, $a(n+1)=a(n)$, and $\beta(n+1)=\beta(n)$. 
    \item[$\blacksquare$] For $j=N$, $N_N=Z_{N}(z)Z_{N \sm 1}(\sm x) Z_{N+1}(\sm x)$
    \begin{align*}
    \mathcal{L}_{\bm \sigma}N_N \ket{G}_{V(N+1)}&=\mathcal{L}(W_{N}(1,1),b) CZ_{N \rightarrow N+1}  \mathcal{L}(W_{\sigma_1}(1,1), b) \cdots \mathcal{L}(W_{\sigma_n}(1,1),b) N_N \ket{G}_{V(N)}\ket{+}_{N+1}, 
    \end{align*}
    where we use the (IH) only for the term $Z_{N \sm 1}(\sm x)$, and using the update rules one finds that $Z_{N \sm 1}(\sm x)$ with the measurements $\mathcal{L}(W_{\sigma_1}(1,1), b) \cdots \mathcal{L}(W_{\sigma_n}(1,1),b)$ is updated to $Z_1(\sm x) Z_N(\sm \beta(n)x)$ such that 
    \begin{align*}
    \mathcal{L}_{\bm \sigma}N_N \ket{G}_{V(N+1)}&\overset{\text{(IH)}}{=} \mathcal{L}(W_{N}(1,1),b) CZ_{N \rightarrow N+1} Z_1(\sm x)Z_{N}(z\sm  \beta(n)x) Z_{N+1}(\sm x) \ket{G'}_{V(N)} \ket{+}_{N+1}\\
    &\uprule Z_1(z\sm (1+\beta(n))x)Z_{N+1}(z\sm (1+\beta(n))x) \mathcal{L}_{\bm \sigma}\ket{G}_{V(N+1)},
    \end{align*}
    where to prove (IS) we identify $\alpha(n+1)=\beta(n+1)=1+\beta(n)$ and $a(n+1)=1/(1+\beta(n))$.
    \end{itemize}
    \item \textbf{Case 2. }$\mathcal{L}(W_{N \sm 1}(1,1),b) $ acts before $\mathcal{L}(W_{N}(1,1),b) $ such that $\mathcal{L}(W_{N}(1,1),b) $ commutes through to the last acting term since it commutes with all other measurement operators, and we return to \textbf{Case 1.}
    \item \textbf{Case 3. }$\mathcal{L}(W_{N \sm 1}(1,1),b) $ acts after $\mathcal{L}(W_{N}(1,1),b) $ so we commute $\mathcal{L}(W_{N}(1,1),b) $ through to the first acting term and get 
    \begin{equation*}
    \mathcal{L}_{\bm \sigma}N_j \ket{G}=\mathcal{L}(W_{\sigma_1}(1,1),b) \cdots \mathcal{L}(W_{\sigma_{n}}(1,1),b) \mathcal{L}(W_{N}(1,1),b)  N_j \ket{G}.
    \end{equation*}
    Depending on which qudit $j\in\{2, \dots, N\}$ the noise operator $N_j$ is acting on, we get the following three cases:
    \begin{itemize}
        \item[$\blacksquare$] For $j<N \sm 1$, $N_j$ and $\mathcal{L}(W_{N}(1,1),b)$ commute thus,
        \begin{align*}
        \mathcal{L}_{\bm \sigma}N_j \ket{G}&=\mathcal{L}(W_{\sigma_1}(1,1),b) \cdots \mathcal{L}(W_{\sigma_n}(1,1),b) N_j \mathcal{L}(W_{N}(1,1),b) \ket{G} \\
        &\overset{\text{(IH)}}{=} Z_1(\alpha(n)(a(n)z \sm x))Z_{N}(\beta(n)(a(n)z \sm x)) \mathcal{L}_{\bm \sigma}\ket{G},
        \end{align*}
        where to prove (IS) we identify $n=n+1$.
        \item[$\blacksquare$] For $j=N \sm 1$: 
        \begin{align*}
        \mathcal{L}_{\bm \sigma}N_{N \sm 1} \ket{G} &\uprule \mathcal{L}(W_{\sigma_1}(1,1),b) \cdots \mathcal{L}(W_{\sigma_n}(1,1),b)  Z_{N\sm 1}((z \sm x))Z_{N \sm 2}(\sm x)Z_{N}(\sm x) \mathcal{L}(W_{N}(1,1),b) \ket{G} \\
        &\overset{\text{(IH)}}{=} Z_1(\alpha(n)(a(n)(z \sm x)\sm x))Z_{N+1}(\beta(n)(a(n)(z \sm x) \sm x)) \mathcal{L}_{\bm \sigma}\ket{G}\\
        &= Z_1\left(\alpha(n)(a(n)+1)\left(\frac{za(n)}{a(n)+1} \sm x\right)\right) Z_{N+1}\left(\beta(n)(a(n)+1)\left(\frac{za(n)}{a(n)+1} \sm x\right)\right)\mathcal{L}_{\bm \sigma}\ket{G},
        \end{align*}
        where to prove (IS) we define $\alpha(n+1)=\alpha(n)(a(n)+1)$, $\beta(n+1)=\beta(n)(a(n)+1)$, and $ a(n+1)=a(n)/(a(n)+1) $.
        \item[$\blacksquare$] For $j=N$, $N_N=Z_{N}(z)Z_{N \sm 1}(\sm x) Z_{N+1}(\sm x)$:
        \begin{align*}
        \mathcal{L}_{\bm \sigma}N_{N} \ket{G}&\uprule \mathcal{L}(W_{\sigma_1}(1,1),b) \cdots \mathcal{L}(W_{\sigma_n}(1,1),b)  Z_{N \sm 1}(z \sm x) Z_{N + 1}(z \sm x) \mathcal{L}(W_{N}(1,1),b)\ket{G} \\
        &\overset{\text{(IH)}}{=} Z_1(\alpha(n)a(n)(z \sm x) ) Z_{N+1}(\beta(n)a(n)(z \sm x))Z_{N+1}(z \sm x)\mathcal{L}_{\bm \sigma}\ket{G} \\
        &=Z_1(\alpha(n)a(n)(z \sm x) ) Z_{N+1}((\beta(n)a(n)+1)(z \sm x))\mathcal{L}_{\bm \sigma}\ket{G},
        \end{align*}
        where to prove (IS) we define $\alpha(n+1)=\alpha(n) a(n)$, $\beta(n+1)=\beta(n)a(n)+1$, and $ a(n+1)=1 $.
    \end{itemize}
\end{itemize}

Therefore, we have proven our induction hypothesis (IH), and if we define $u=a(n)z \sm x$ and we identify $\alpha=\alpha(n)$ and $\beta=\beta(n)$, then we obtain that for $j\in\{2, \dots N \sm 1\}$ being one of the middle qudits
\begin{equation*}
    \mathcal{L}_{\bm\sigma}N_j\ket{G} = Z_1(\alpha u) Z_N(\beta u) \mathcal{L}_{\bm\sigma}\ket{G},
\end{equation*}
which proves Eq.~\eqref{eq.minner}. 

\subsection{Fidelity of the generalized Bell pair} \label{SubSec.Fidelity of final state}
The fidelity of the final generalized Bell pair is  $F=\bra{G'}\widetilde{\mathcal{E}}_N\cdots\widetilde{\mathcal{E}}_1\rho'\ket{G'}$, where $\widetilde{\mathcal{E}}_i$ are the updated noise maps following the descriptions of Eq.~\eqref{eq.mtarget} and Eq.~\eqref{eq.minner}, and $\rho'=\ketbra{G'}$ with $\ket{G'}$ being the generalized Bell pair between qudits 1 and $N$. Since the final state is small, we derive an analytical formula for the fidelity of the final state for a given weight vector $\bm w$ that encodes the measurement strategy of the linear cluster. We make use of the fact that applying $w$ times a map of the shape of Eq.~\eqref{eq.mtarget} or Eq.~\eqref{eq.minner} with parameter $\lambda$ is the same as applying it once with parameter $\lambda^w$. In order to derive the fidelity formula, we take the following six steps:
\begin{enumerate}[label=\textbf{\arabic* -}]
\item\textbf{Composition of two middle qudit maps. }First, we compute the action of two different maps of the middle qudits, Eq.~\eqref{eq.minner}, with noise parameters of the shape $\lambda^w$, where $w$ is different for each map.
\begin{align*}
\mathcal{M}(a,b)^y \mathcal{M}(i,j)^x \rho' &=\mathcal{M}(a,b)^y \left( \lambda^x \rho' + \frac{1 \sm \lambda^x}{d}  \sum_{u \in \Fd} Z_1(i u) Z_N(j u) \rho' (Z_1(i u) Z_N(j u))^\dag \right) \\
&= \lambda^{y+x} \rho' + \frac{\lambda^y \sm \lambda^{x+y}}{d}\sum_{u \in \Fd} Z_1(iu) Z_N(ju) \rho' (Z_1(iu)Z_N(ju))^{\dag} \\
&+ \frac{\lambda^x \sm \lambda^{y+x}}{d}\sum_{k \in \Fd} Z_1(ka)Z_N(kb) \rho' (Z_1(ka)Z_N(kb))^{\dag} \\
&+ \frac{(1 \sm \lambda^x)(1 \sm \lambda^y)}{d^2}\sum_{k,u \in \Fd} Z_1(iu+ka)Z_N(ju+kb) \rho'  (Z_1(iu+ka)Z_N(ju+kb))^{\dag},
\end{align*}
where we can write the last noise term as $\sum_{r,s \in \Fd}Z_1(r) Z_N(s) \rho' (Z_1(r) Z_N(s))^{\dag}$.

\item\textbf{Composition of three middle qudit maps. }Here we compute the action of three different maps of the middle qudits, Eq.~\eqref{eq.minner}, that have noise parameters of the shape $\lambda^w$, where $w$ is different for each map,
\begin{align*}
\mathcal{M}(\alpha,\beta)^z \mathcal{M}(a,b)^y \mathcal{M}(i,j)^x \rho'&= \lambda^{x+y+z} \rho' + \frac{\lambda^z \lambda^y(1 \sm \lambda^x)}{d} \sum_{u \in \Fd} Z_1(iu) Z_N(ju) \rho' (Z_1(iu) Z_N(ju))^{\dag} \\
&+ \frac{\lambda^z \lambda^x(1 \sm \lambda^y)}{d}\sum_{k \in \Fd} Z_1(ka)Z_N(kb) \rho' (Z_1(ka)Z_N(kb))^{\dag} \\
&+ \frac{\lambda^z(1 \sm\lambda^x)(1 \sm\lambda^y)}{d^2}\sum_{r,s \in \Fd} Z_1(r) Z_N(s) \rho' (Z_1(r) Z_N(s))^{\dag} \\
&+ \frac{(1-\lambda^z)\lambda^{y+x}}{d} \sum_{l \in \Fd} Z_1(l \alpha) Z_N(l \beta) \rho' (Z_1(l \alpha) Z_N(l \beta) )^{\dag}\\
&+\frac{\lambda^y(1 \sm \lambda^z)(1 \sm \lambda^x)}{d^2}\sum_{r,s \in \Fd} Z_1(r) Z_N(s) \rho' (Z_1(r) Z_N(s))^{\dag} \\
&+ \frac{\lambda^x(1 \sm \lambda^y)(1 \sm \lambda^z)}{d^2}\sum_{r,s \in \Fd} Z_1(r) Z_N(s) \rho' (Z_1(r) Z_N(s))^{\dag}\\
&+ \frac{(1 \sm \lambda^x)(1 \sm \lambda^y)(1 \sm \lambda^z)}{d^2}\sum_{r,s \in \Fd} Z_1(r) Z_N(s) \rho' (Z_1(r) Z_N(s))^{\dag}.
\end{align*}
We introduce the following abbreviations 
\begin{equation*}
\Sigma(\alpha,\beta):= \sum_{l \in \Fd} Z_1(l \alpha) Z_N(l \beta) \rho' (Z_1(l \alpha) Z_N(l \beta) )^{\dag} \qquad \text{and}\qquad \Sigma:= \sum_{r,s \in \Fd} Z_1(r) Z_N(s) \rho' (Z_1(r) Z_N(s))^{\dag}
\end{equation*}
such that we now have
\begin{align*}
\mathcal{M}(\alpha,\beta)^z \mathcal{M}(a,b)^y \mathcal{M}(i,j)^x \rho' 
&= \lambda^{x+y+z}\rho'\\
&+\frac{\lambda^z\lambda^y(1 \sm \lambda^x)}{d} \Sigma(i,j) + \frac{\lambda^z \lambda^x (1 \sm \lambda^y)}{d} \Sigma(a,b) + \frac{(1 \sm \lambda^z)\lambda^x \lambda^y}{d} \Sigma(\alpha,\beta) \\
&+ \frac{\lambda^z(1 \sm \lambda^x)(1 \sm \lambda^y)+\lambda^y(1 \sm \lambda^z)(1 \sm \lambda^x)+\lambda^x(1 \sm \lambda^y)(1 \sm\lambda^z)}{d^2}\Sigma \\
& + \frac{(1 \sm \lambda^x)(1 \sm \lambda^y)(1 \sm \lambda^z)}{d^2}\Sigma.
\end{align*}

\item\textbf{Composition of four middle qudit maps. }We compute the action of four different maps of the middle qudits, Eq.~\eqref{eq.minner}, that have noise parameters of the shape $\lambda^w$, where $w$ is different for each map,
\begin{align*}
\mathcal{M}(\gamma,\delta)^w \mathcal{M}(\alpha,\beta)^z \mathcal{M}(a,b)^y \mathcal{M}(i,j)^x \rho'
&= \lambda^{x+y+z+w} \rho' \\
&+ \frac{\lambda^w\lambda^z\lambda^y(1 \sm \lambda^x)}{d}\Sigma(i,j)+\frac{\lambda^w\lambda^z\lambda^x(1 \sm \lambda^y)}{d}\Sigma(a,b) \\
&+ \frac{\lambda^w\lambda^x\lambda^y(1 \sm \lambda^z)}{d} \Sigma(\alpha,\beta) + \frac{\lambda^x\lambda^y\lambda^z(1 \sm \lambda^w)}{d} \Sigma(\gamma,\delta)  \\
&+ \frac{\lambda^z(1 \sm \lambda^x)(1 \sm \lambda^y)(1 \sm \lambda^w)+\lambda^y(1 \sm \lambda^z)(1 \sm \lambda^x)(1 \sm \lambda^w)}{d^2}\Sigma  \\
&+ \frac{\lambda^x(1 \sm \lambda^y)(1 \sm \lambda^z)(1 \sm \lambda^w)+\lambda^w(1 \sm \lambda^x)(1 \sm \lambda^y)(1 \sm \lambda^z)}{d^2}\Sigma \\
&+ \frac{\lambda^w\lambda^z(1 \sm \lambda^x)(1 \sm \lambda^y)+\lambda^y\lambda^w (1 \sm \lambda^z)(1 \sm \lambda^x)}{d^2} \Sigma \\
&+ \frac{\lambda^x\lambda^w(1 \sm \lambda^y)(1 \sm \lambda^z) +(1 \sm \lambda^x)(1 \sm \lambda^y)(1 \sm \lambda^z)(1 \sm \lambda^w)}{d^2} \Sigma.
\end{align*}

\item\textbf{Composition of $n$ middle qudit maps. }Using the results from the previous steps, we can generalize the composition of $n$ different maps of the middle qudits, Eq.~\eqref{eq.minner}, that have noise parameters of the shape $\lambda^w$, where $w$ is different for each map. First, we introduce some definitions:
\begin{align*}
I=\left((x_i,a_i,b_i) \right)^H_{i=1}, \qquad\text{and}\qquad \bar{\mathcal{M}}(x,a,b):=\mathcal{M}(a,b)^x
\end{align*}
such that we can write $\bar{\mathcal{M}}(I(i))$. Then, we define a function that encodes the noise parameters that accompany the $\Sigma$ term of the composite noise map as
\begin{equation*}
    \pi_h(\pmb x,\lambda,I):=\sum_{(i^{j}_{k_1},\dots,i^{j}_{k_h}) \in U_h} \prod^h_{l=1}(1 \sm \lambda^{x_{i^{j}_{k_l}}})\prod_{x_i \in I \setminus \{x_{i^{j}_{k_1}},\dots,x_{i^{j}_{k_h}} \}} \lambda^{x_i},
\end{equation*}
where
\begin{align*}
U_h:= \{ (i^{j}_{k_1},\dots,i^{j}_{k_h}) | \exists i^{j}_{k_r} \in (i^{j}_{k_1},\dots,i^{j}_{k_h})| i^{j}_{k_r} \not \in (i^{m}_{k_1},\dots,i^{m}_{k_h}),  \ \forall m \neq j \}.
\end{align*}
Inductively, we see that 
\begin{align*}
\prod^H_{i=1} \bar{\mathcal{M}}(I(i))\rho'= \lambda^{\sum_i x_i}\rho'+\frac{1}{d}\sum_{i}(1 \sm \lambda^x_i)\lambda^{\sum_{j \neq i} x_j} \Sigma(a_i,b_i)+ \frac{\sum^H_{h=2}\pi_h(\pmb x,\lambda,I)}{d^2}\Sigma.
\end{align*}
Now we set $x_i=w_i$ such that they represent the elements of the weight vector $\bm w$ and find that the action of the composition of all $n$ updated noise maps of the middle qudits is 
\begin{align*}
\widetilde{\mathcal{E}}_2\cdots \widetilde{\mathcal{E}}_{N \sm 1}\rho'=\prod^{p^2}_{i=1} \bar{\mathcal{M}}(I(i))\rho'= \lambda^{\sum_i w_i}\rho'+\frac{1}{d}\sum_{i}(1 \sm \lambda^{w_i})\lambda^{\sum_{j \neq i} w_j} \Sigma(a_i,b_i)+ \frac{\sum^{p^2}_{h=2}\pi_h(\pmb w,\lambda,I)}{d^2}\Sigma.
\end{align*} 
\item\textbf{Composition of all qudit maps.} We take the result from the previous step and add the updated noise maps of the edge qudits, 1 and $N$. These are presented in Eq.~\eqref{eq.mtarget}, and are such that
\begin{align*}
\widetilde{\mathcal{E}}_1\widetilde{\mathcal{E}}_N \rho'= \lambda^2 \rho' + \frac{1 \sm \lambda^2}{d^2} \Sigma.
\end{align*}
Therefore, the final noisy generalized Bell pair is
\begin{align*}
\widetilde{\mathcal{E}}_1\cdots \widetilde{\mathcal{E}}_N\rho'&=\lambda^{2+\sum_{i=1}^{p^2}w_i}\rho'+ \frac{1}{d}
\sum_{i=1}^{p^2}(1 \sm \lambda^{w_i}) \lambda^{2+\sum_{j \neq i}^{p^2}w_j}\Sigma(a_i,b_i) \\
&+\frac{1}{d^2} \left(\sum_{h=2}^{p^2}\pi_h(\pmb w,\lambda,I)+\lambda^{\sum_{i=1}^{p^2}w_i}(1 \sm \lambda^2)+\sum_{i=1}^{p^2}(1 \sm \lambda^2)(1 \sm \lambda^{w_i})\lambda^{\sum_{j \neq i} w_j} \right) \Sigma.
\end{align*}

\item\textbf{Compute fidelity. }We compute the formula for the fidelity using the final composite noise map such that only terms of the composite map that are proportional to the identity contribute to the fidelity, yielding
\begin{align*}
F&=\lambda^{2+|\pmb w|}+\frac{1}{d}\left(\sum^{p^2}_{i=1}(1 \sm \lambda^{w_i})\lambda^{2+\sum_{j \neq i}w_j }   \right) \\
&+ \frac{1}{d^2} \left(\sum^{p^2}_{h=2} \pi_h(\pmb w, f, I) + \lambda^{|\pmb w|}(1 \sm \lambda^2)+\sum^{p^2}_{i=1}(1 \sm \lambda^2)(1 \sm \lambda^{w_i})\lambda^{\sum_{j \neq i} w_j})    \right),
\end{align*}
where the fidelity depends on the weight vector $\bm w$, the dimension $d=p^m$, and the depolarizing parameter $\lambda$. 
\end{enumerate}
\subsection{Side-to-side strategy}\label{SubSec.Side to side strategy}
Here we derive the weight vector for the side-to-side strategy. Given a general noise term from the depolarizing channel, Eq.~\eqref{Eq.Depol}, $N_j=Z_j(z)Z(\sm A_j x)$ and the side-to-side $\mathcal{L}_{\bm \sigma}$, i.e., $\bm\sigma = (2, 3, \dots, N \sm 1)$, depending on which qudit $j$ the noise operator acts on we discuss three different cases. Below we denote $\ket{G^*}$ for a linear chain of reduced size obtained from a part of the measurement procedure $\mathcal{L}_{\bm \sigma}$.
\begin{itemize}
    \item \textbf{Case 1. }For $j\in (3,N \sm 2)$ the noise operator is $N_j=Z_{j+1}(\sm x)Z_j(z) Z_{j\sm 1}(\sm x)$
    \begin{align*}
    \mathcal{L}_{\bm\sigma}N_j\ket{G}&=\mathcal{L}(W_{2}(1,1),b) \cdots \mathcal{L}(W_{j+1}(1,1),b)Z_{j+1}(\sm x)Z_j(z) Z_{j\sm 1}(\sm x) \ket{G^*} \\
    &\uprule \mathcal{L}(W_{2}(1,1),b) \cdots \mathcal{L}(W_{j}(1,1),b) Z_N(\sm x) Z_j(z\sm x)Z_{j\sm 1}(\sm x) \ket{G^*} \\
    &\comrel \mathcal{L}(W_{2}(1,1),b) \cdots \mathcal{L}(W_{j\sm 1}(1,1),b) Z_N(z\sm 2x) Z_{j\sm 1}(z\sm 2x) \ket{G^*} \\
    &\uprule \mathcal{L}(W_{2}(1,1),b) \cdots \mathcal{L}(W_{j\sm 2}(1,1),b) Z_N(2(z\sm 2x)) Z_{j \sm 2}(z\sm 2x) \ket{G^*},
    \end{align*} 
    and then inductively we get $\mathcal{L}_{\bm\sigma}N_j\ket{G}= Z_N((j\sm 1)(z\sm 2x))Z_1(z\sm 2x) \mathcal{L}_{\bm\sigma}\ket{G}$ such that the updated noise map is $\widetilde{\mathcal{E}}_j=\mathcal{M}(1, j \sm 1)$.
    \item \textbf{Case 2. } For $j=N \sm 1$ the noise operator is $N_{N \sm 1}=Z_{N}(\sm x)Z_{N \sm 1}(z) Z_{N \sm 2}(\sm x)$,
    \begin{align*}
    \mathcal{L}_{\bm\sigma}N_{N \sm 1}\ket{G} &\uprule \mathcal{L}(W_{2}(1,1),b) \dots \mathcal{L}(W_{N \sm 2}(1,1),b) Z_{N}(z\sm x) Z_{N\sm 2}(z\sm x) \ket{G^*} \\
    &\uprule \mathcal{L}(W_{2}(1,1),b) \dots \mathcal{L}(W_{N \sm 3}(1,1),b) Z_{N}(2(z\sm x)) Z_{N\sm 3}(z\sm x) \ket{G^*} \\
    &\uprule \cdots \uprule Z_1(z\sm x)Z_N((N\sm 2)(z\sm x)) \mathcal{L}_{\bm\sigma}\ket{G}
    \end{align*}
    such that the updated noise map is $\widetilde{\mathcal{E}}_{N\sm 1}=\mathcal{M}(1, N\sm 2)$.
    \item \textbf{Case 3. }For $j=2$ the noise operator is $N_2=Z_{1}(\sm x)Z_{2}(z) Z_{3}(\sm x) $
    \begin{align*}
    \mathcal{L}_{\bm\sigma}N_2\ket{G}&= \mathcal{L}(W_{2}(1,1),b) \mathcal{L}(W_3(1,1),b) Z_{1}(\sm x)Z_{2}(z) Z_{3}(\sm x) \ket{G^*} \\
    &\uprule  \mathcal{L}(W_{2}(1,1),b) Z_{1}(\sm x)Z_{2}(z \sm x) Z_N(\sm x) \ket{G^*}  \\
    &= Z_1(z \sm x)Z_N(z \sm x)  \mathcal{L}_{\bm\sigma}\ket{G}
    \end{align*}
    such that the updated noise map is $\widetilde{\mathcal{E}}_2=\mathcal{M}(1, 1)$.
\end{itemize}

Therefore, we have $\widetilde{\mathcal{E}}_j= \mathcal{M}(1,j \sm 1)$. The weight of map $\mathcal{M}(\alpha,\beta)$ is $w^{\beta}_{\alpha}$, where $\alpha$ and $\beta$ are from $\Fp$, so we have to compute their effect as a weight vector entry modulo $p$. Note that the weight vector $\bm w$ captures the number of times the updated noise maps $\mathcal{M}(\alpha,\beta)$ are applied to the generalized Bell pair. The first $p$ entries of the weight vector $w_0^0,\dots,w_0^{p \sm 1}$ capture the number of times the maps $\mathcal{M}(0,0),\dots, \mathcal{M}(0,p \sm 1)$ act on the final state. For the side-to-side strategy, the maps $\mathcal{M}(0,0),\dots, \mathcal{M}(0,p \sm 1)$ are not applied onto the target state, so the first $p$ weights are $0$. The second $p$ entries of the weight vector $w_1^0,\dots,w_1^{p \sm 1}$ capture the number of times the maps $\mathcal{M}(1,0),\dots, \mathcal{M}(1,p \sm 1)$ act on the final state. The side-to-side strategy has the maps $\{\mathcal{M}(1,k \mod p)\}_{k=1}^{n}$ acting on the final state, where we note that the maps where $k$ differs by $p$ correspond to the same map. Therefore, dividing the number of middle qudits $n$ by $p$ gives $n=mp+s$, where $m \in \mathbbm{N}_0$ is the number of times $p$ divides $n$ and $s < p$ is the remainder of the division. Thus, $m$ captures the number of times all maps of the form $\mathcal{M}(1,0),\dots, \mathcal{M}(1,p \sm 1)$ are applied to the generalized Bell pair, while $s$ captures the additional maps $\mathcal{M}(1,1),\dots, \mathcal{M}(1,s)$ applied to the final state. In total this results in the weight vector elements $w_1^0=m, w_1^1=m+H(1 \sm s), \dots, w_1^{p \sm 1}+H(p \sm 1 \sm s)$, where $H(x)=1$ for $x \leq 0$ and $H(x)=0$ for $x > 0$. Similarly, the weight vector elements $w_l^0, \dots, w_l^{p \sm 1}$ capture the number of times the maps $\mathcal{M}(l,0), \dots \mathcal{M}(l,p \sm 1)$ are applied to the final state, for $l=2, \dots , p \sm 1$, which are not applied in the case of the side-to-side strategy, so the corresponding weights are $0$.  Thus, this gives us the weight vector presented in Eq.~\eqref{eq:sidetoside}. 

\section{Further results} \label{Sec.Physically motivated noise models}
In Sec.~\ref{Sec.Application of method}, we analyze the manipulation of an open linear cluster subject to two noise sources into a generalized Bell pair between the ends of the linear cluster. The two considered noise sources are modeled by single-qudit depolarizing noise, Eq~\eqref{Eq.Depol}, with different noise parameters $1 \sm \lambda$, where one has a dimension-independent parameter $\lambda=r$ and the other has a dimension-dependent one $\lambda=q_d$. 

The dimension-dependent parameter considered in Sec.~\ref{Sec.Application of method} is defined by setting the Choi-Jamio\l kowski fidelity of $m$ qubit depolarizing channels, each with $\lambda=q_2$, to match the Choi-Jamio\l kowski fidelity of one $2^m$ dimensional depolarizing channel, with $\lambda=q_d$. This has the consequence that the critical depolarizing probability from the qubit depolarizing channel, i.e., the depolarizing probability such that a generalized Bell pair is still entangled, is mapped onto the critical depolarizing probability of the $2^m$-dimensional depolarizing channel. The critical depolarizing parameter for dimension $d$ is given by $q_d^{\text{crit}}=1/(d+1)$, which can be obtained via equating the critical fidelity for dimension $d$, given by $F_d^{\text{crit}}=1/d$, with the Choi-Jamio\l kowski fidelity of the qudit depolarizing channel. By inserting the qubit critical probability into the adaptation formula, Eq.~\eqref{Equ:tildef}, one can easily see that the critical depolarizing probabilities match.

Besides the dimensional dependency studied in Sec.~\ref{Sec.Application of method}, here we explore other scenarios motivated by the experimental realization of high-dimensional systems in trapped ions, specifically in Ref.~\cite{ringbauer2022universal}. In the supplementary information in Ref.~\cite{ringbauer2022universal}, two error rate dependencies are described. The first type of error is the read-out error, where the full $d$-dimensional read-out is performed as a sequence of linearly many two-level read-out procedures, resulting in the dimension-dependent depolarizing parameter modeled as $q_d = q_2^{\,d/2}$. The second error type is the single-qudit gate error, where each single-qudit gate is decomposed into $d(d \sm 1)/2$ two-level rotations, and, thus, we model the dimension-dependent depolarizing parameter as $q_d = q_2^{\,d(d  \sm  1)/2}$.

\begin{figure}[h]
    \centering
    \subfloat[]{\label{fig:fidlinear}\includegraphics[width=0.5\columnwidth]{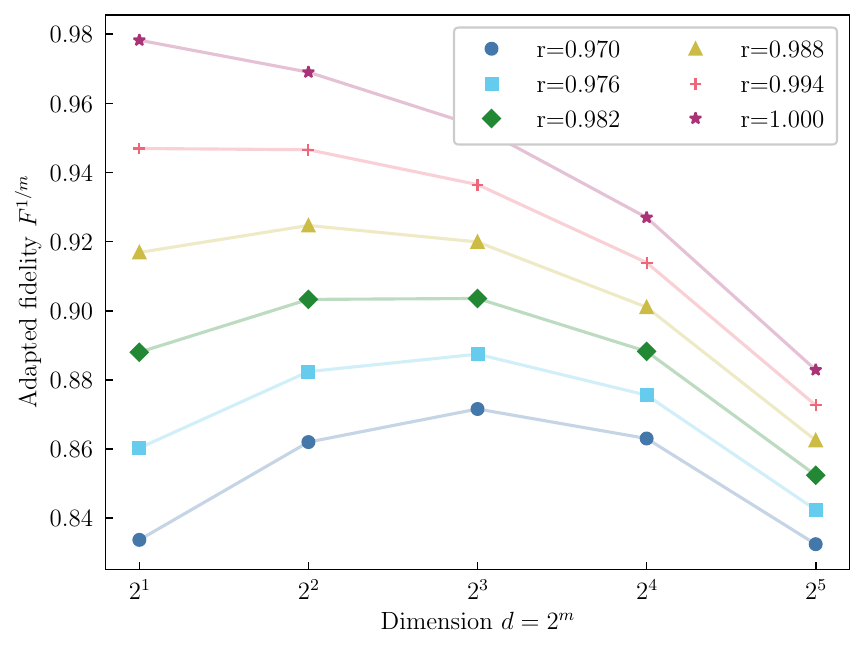}} 
    \subfloat[]{\label{fig:fidquadratic}\includegraphics[width=0.5\columnwidth]{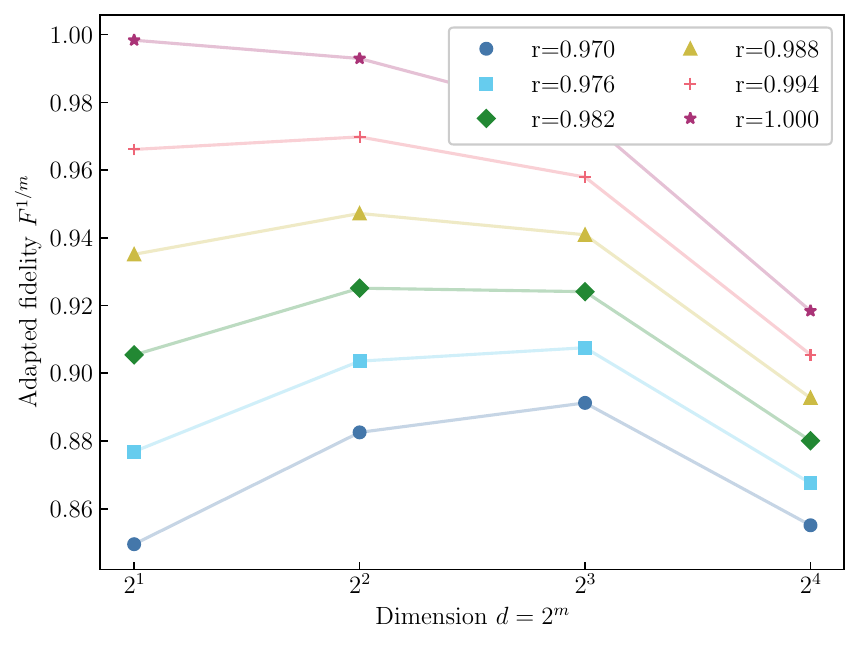}}
    \caption{Adapted fidelity $F^{1/m}$ of a generalized Bell pair obtained via the side-to-side measurement strategy on a noisy linear cluster of size $N=10$ in terms of the local dimension $d=2^m$, for different dimension-independent depolarizing parameters $1 \sm r$. The dimension-dependent depolarizing parameter, $1 \sm q_d$, is (a) linear, such that $q_d=q_2^{d/2}$ with $q_2=0.996$, and (b) quadratic, such that $q_d=q_2^{d(d\sm 1)/2}$ with $q_2=0.9997$.}
    \label{fig:fidelity_phys}
\end{figure}

As in Sec.~\ref{Sec.Application of method}, we concatenate again a dimension-dependent depolarizing noise with a dimension-independent one, governed by $\lambda=r$, and we apply it to a linear cluster which is manipulated with the side-to-side strategy to achieve a generalized noisy Bell pair. Using the formula of the fidelity given in Appendix~\ref{Sec.Proofs: Linear chain} and the weight vector of the side-to-side strategy, Eq.~\eqref{eq:sidetoside}, we compute the adapted fidelity $F^{1/m}$ in terms of the local dimension $d=p^m$, which is presented in Fig.~\ref{fig:fidelity_phys}. There one can observe that in certain parameter regimes intermediate qudit dimensions yield the optimal performance.

\end{document}